\documentclass[useAMS,usenatbib]{mn2e}
\usepackage{url}
\usepackage{color}
\usepackage{pdfpages}
\usepackage[colorlinks=true,citecolor=blue]{hyperref}
\usepackage{graphicx,graphics,color}
\usepackage{amsmath}
\usepackage{soul}
\usepackage{amssymb}
\usepackage[normalem]{ulem}

\newcommand\chandra{{\it Chandra~}}

\newcommand\kms{\ifmmode {\rm~km\ s}^{-1} \else ~km s$^{-1}$\fi}
\newcommand\Hunit{\ifmmode {\rm~km\ s}^{-1}\ {\rm Mpc}^{-1}
        \else ~km s$^{-1}$ Mpc$^{-1}$\fi}
\newcommand\ctssec{\ifmmode {\rm~count\ s}^{-1} \else ~count s$^{-1}$\fi}
\newcommand\ergsec{\ifmmode {\rm~erg\ s}^{-1} \else
        ~erg s$^{-1}$\fi}
\newcommand\ergs{\ifmmode {\rm~erg\ s}^{-1} \else
        ~erg s$^{-1}$\fi}
\newcommand\funit{\ifmmode {\rm~erg\ s}^{-1}\;{\rm cm}^{-2} \else
        ~ergs s$^{-1}$ cm$^{-2}$\fi}
\newcommand\phflux{\ifmmode {\rm~photon\ s}^{-1}\;{\rm cm}^{-2}
        \else   ~photon s$^{-1}$ cm$^{-2}$\fi}
\newcommand\efluxA{\ifmmode {\rm~erg\ s}^{-1}\;{\rm cm}^{-2}\;{\rm
        \AA}^{-1} \else ~erg s$^{-1}$ cm$^{-2}$ \AA$^{-1}$\fi}
\newcommand\efluxHz{\ifmmode {\rm~erg\ s}^{-1}\;{\rm cm}^{-2}\;{\rm
        Hz}^{-1} \else ~erg s$^{-1}$ cm$^{-2}$ Hz$^{-1}$\fi}
\newcommand\cc{\ifmmode {\rm~cm}^{-3} \else cm$^{-3}$\fi}
\newcommand\FWHM{\ifmmode {\rm~FWHM} \else ${\rm~FWHM}$\fi}
\newcommand\Msun{\ifmmode M_{\odot} \else $M_{\odot}$\fi}
\newcommand\Zsun{\ifmmode Z_{\odot} \else $Z_{\odot}$\fi}
\newcommand\Lsun{\ifmmode L_{\odot} \else $L_{\odot}$\fi}

\newcommand\hbeta{\ifmmode {\rm H}\beta \else H$\beta$\fi}
\newcommand\Kalpha{\ifmmode {\rm K}\alpha \else K$\alpha$\fi}
\newcommand\nh{\ifmmode N_{\rm H} \else N$_{\rm H}$\fi}

\newcommand{\mnras}{MNRAS}

\def\H2{\hbox{H$_{2}$}}

\title[Cold front in Abell~2626]{Merging cold front and AGN feedback in the peculiar galaxy cluster Abell~2626}

\author[Kadam et al.]{S. K. Kadam$^{1}$, S. S. Sonkamble$^{2}$, P. K. Pawar$^{3}$, M. K. Patil$^{1}$\thanks{\color{blue}{patil@associates.iucaa.in}} \\ \\
$^{1}$School of Physical Sciences, Swami Ramanand Teerth Marathwada University, Nanded-431 606, India\\
$^{2}$National Centre for Radio Astrophysics (NCRA), Tata Institute of Fundamental Research (TIFR), Pune-411 007, India\\
$^{3}$Inter-University Centre for Astronomy and Astrophysics (IUCAA), Pune-411 007, India\\
}

\begin{document}
\pagerange{\pageref{firstpage}--\pageref{lastpage}} \pubyear{2017}
\maketitle
\label{firstpage}
\begin{abstract}

This paper presents the analysis of a combined 134 ks {\it Chandra} data of a peculiar galaxy cluster Abell 2626. This study confirms the earlier detection of the east cavity at $\sim$13 kpc and reports detection of a new cavity at $\sim$39 kpc on the west of the X-ray peak. The average mechanical power injected by the AGN outburst ${\rm P_{cav} \sim 6.6 \times 10^{44}\, erg\, s^{-1}}$ is $\sim$29 times more than required to compensate the cooling luminosity ${\rm L_{cool} = 2.30 \pm 0.02 \times 10^{43} \ergs}$. The edges in the SB on the west and south-west at $\sim$36 kpc and 33 kpc, respectively, have the gas compressions of 1.57$\pm$0.08 and 2.06$\pm$0.44 and are spatially associated with the arcs in the temperature and metallicity maps due to the merging cold fronts. The systematic study of the nuclear sources exhibited dramatic changes over the span of ten years. The NE source that emitted mostly in the soft band in the past disappeared in the recent observations. Instead, an excess emission was seen at $2.2''$ on its west and required an unrealistic line of sight velocity of $\sim$ $675\times{}c$ if is due to its movement. The count rate analysis and spectral analysis exhibited a change in the state of the SW source from a soft state to the hard due to the change in the mass accretion rate. No such spectral change was noticed for the NE source. \\
\end{abstract}


\section[1]{Introduction}

The copious amount of X-ray emission from galaxy clusters require that a large fraction of the intracluster medium (ICM) must lose its thermal energy through the  bremsstrahlung radiation. As a result, in the absence of any heating source, to support the weight of the overlaying layers, a cooling flow is expected to initiate at the core of such clusters \citep{1994ARA&A..32..277F}. The inferred rates of mass depositions onto the cores of such clusters using X-ray observations would be several hundreds to thousands of solar mass a year \citep[for a review see][]{1994ARA&A..32..277F}. Though a wealth of evidences supporting condensation of the ICM in the cores of cooling flow clusters are available, the amount of cool gas detected in reservoirs is a small fraction of that predicted by the standard cooling flow model \citep{2006PhR...427....1P,2007ARA&A..45..117M,2012NJPh...14e5023M}. The precise high resolution XMM\,RGS spectroscopic observations of X-ray emitting gas from cooling flow clusters failed to exhibit signatures of the gas that has cooled below 1-2 keV \citep{2003ApJ...590..207P}. Further, spectroscopic and imaging studies of the cores of such clusters in UV and optical bands inferred the star formation rates significantly lower than predicted by the model \citep{1987MNRAS.224...75J}. These discrepancies between the observed and the predicted properties led to the \textit{cooling flow problem} and invited the necessity of a powerful heating at the core to resolve this issue.
Several heating mechanisms have been proposed to suppress the over cooling of the ICM, which include the electron thermal conduction \citep{2003ApJ...582..162Z}, cosmic rays \citep{1991ApJ...377..392L} and supernovae \citep{2004ApJ...601..173M}. However, the mechanical heating by Active Galactic Nucleus (AGN) was regarded as the most prevalent solution to the cooling flow problem \citep{1994ARA&A..32..277F,1995MNRAS.276..663B,2002MNRAS.332..729C,2004ApJ...615..681R,2007ARA&A..45..117M, 2012AdAst2012E...6G}. The ``radio-mode" of mechanical heating  assumes a feedback loop, where the cool collapsed ICM accreted onto the nuclear black hole ignites the AGN and in its return develops energetic outbursts. The AGN outbursts deposit sufficiently large amount of energy into the surrounding ICM and hence quench the cooling flows in the cores of such clusters \citep[for reviews see,][]{2007ARA&A..45..117M,2012ARA&A..50..455F,2012AdAst2012E...6G}. Several studies using high resolution X-ray data advocate the viability of this mode of heating \citep[e.g.,][]{2004ApJ...607..800B, 2008MNRAS.385..757D, 2013Ap&SS.345..183P}, however, the exact mode in which the AGN transfers its energy to the ICM is not yet fully understood.

X-ray deficient bubbles or cavities, jets, outflows, shocks, sonic ripples, and sharp density discontinuities in the X-ray surface brightness distribution are some of the most discernible evidences supporting the radio mode of the AGN heating at the core of such clusters \citep{2007PhR...443....1M,2007ARA&A..45..117M,2012AdAst2012E...6G}. Particularly, the amazing details provided by the sub-arcsecond spatial resolution capability of the \chandra telescope has provided us with the clear evidences regarding detection of few to few tens kpc sized under-dense X-ray deficit bubbles or cavities in the wake of the ICM of dozens of galaxy groups and clusters. As a result cavity investigation and analysis has received a systematic attention in the literature over last two decades. Few of the well studied systems includes e.g., Perseus \citep{2003MNRAS.344L..43F, 2006MNRAS.366..417F}, Cygnus A \citep{2006ApJ...644L...9W}, Hydra A \citep{2005ApJ...628..629N}, M87 \citep{2005ApJ...635..894F}, Abell 2052 \citep{2001ApJ...558L..15B, 2009ApJ...697L..95B, 2011ApJ...737...99B}, Abell 2597 \citep{2012MNRAS.424.1026T}, Abell 1991 \citep{2013Ap&SS.345..183P}, Abell 3847 \citep{2017MNRAS.466.2054V}, ZwCl 2701 \citep{2016MNRAS.461.1885V}, Abell 2390 \citep{2015Ap&SS.359...61S}, NGC 5813 \citep{2015ApJ...805..112R}, HCG 62 \citep{2010ApJ...714..758G}, NGC 6338 \citep{2012MNRAS.421..808P}, where the presence of X-ray cavities are well evident. These cavities or bubbles are the X-ray deficit low density regions in the surface brightness distribution, which float outward in the cooling flow atmosphere until reaches equilibrium at some radius where the ambient entropy becomes equal to that within the bubble \citep{2006ApJ...652..216R}. The most accepted explanation for the formation of such cavities is that the radio jets launched by the central AGN carve these cavities, which are then lifted in the ICM \citep{1994ARA&A..32..277F}. The spatial association of the X-ray cavities with the radio jets and lobes in majority of the cooling flow clusters   strongly support this scenario of cavity formation \citep{2012AdAst2012E...6G, 2012ARA&A..50..455F}.\\

\begin{table*}
\caption{Details of \textit{Chandra} X-ray observations of A2626}
\begin{tabular}{@{}ccccccccr@{}}
\hline
{\it ObsID}    &Instrument & Obs Date &Data Mode &PI &Exposure (ks) & Cleaned Exposure (ks) \\
\hline
\textit{3192}  &ACIS-S &22 Jan. 2003 &VFAINT &Sarazin &24.7  &24.3  \\
\textit{16136} &ACIS-S &20 Oct. 2013 &VFAINT &Sarazin &110.8 &110.1 \\
\hline
\end{tabular}
\footnotesize 
\begin{flushleft} 
\end{flushleft}
\label{obs}
\end{table*}

The enthalpy content of X-ray cavities provide a reliable way of quantifying the lower limits on the  energy input by the AGN outburts. Here, the underlying assumption is that the X-ray cavities are in pressure equilibrium with the surrounding gas. Thus, investigation of X-ray cavities and their systematic analysis act as an important probe to assess the role of the AGN feedback in transferring heat to the surrounding ICM \citep{2004ApJ...607..800B, 2006ApJ...652..216R}. Though numerous studies have been conducted using high-resolution \chandra data in this regard, however, we still lack in clear understanding of the exact mechanism of the AGN heating  \citep{2003ApJ...590L...5M}. It is still not clear what fraction of the mechanical power supplied by the AGN goes in ICM heating. Therefore, a systematic study of investigating the correlation between the ICM cooling and the extent of the mechanical power in the form of X-ray cavities in a large sample of galaxy clusters hosting cool cores is of fundamental importance in delineating the impact of the AGN feedback in the cores of galaxy clusters.

This paper is aimed to investigate AGN feedback in the core of a galaxy cluster Abell 2626 (A2626 hereafter) and to examine the possibility of the AGN-regulated heating of the ICM. This paper also assesses appropriateness of the AGN heating with the cooling of the ICM on the basis of the radio power of the central black hole. This has been achieved by analyzing the publicly available two separate \chandra observations of net exposure 134 ks and 1.4 GHz VLA radio data on A2626. A2626 has been identified as a complex merging galaxy cluster \citep{1999ApJS..125...35S} hosting a double nucleus at its core \citep{2008ApJ...682..155W}. This cluster is also reported to host a radio mini-halo at $\sim$70 kpc from its centre \citep{2008ApJ...682..155W} with a giant kite shaped diffuse radio emission formed due to three radio arcs  \citep{2004A&A...417....1G}. In a recent study using low frequency 610 MHz GMRT observations \citet{2017MNRAS.466L..19K} have reported the detection of the fourth radio arc making the radio emission to take the kite shaped morphology. Two cold fronts have also been reported in this cluster associated with the radio arcs \citep{2018A&A...610A..89I}. \cite{2000AJ....119...21R} found an X-ray excess spatially associated with the radio source, however, failed to detetct the X-ray deficit (bubbles) regions formed by the radio plasma. Later, \cite{2016ApJS..227...31S} using 24 ks \chandra data detected one X-ray deficit cavity. As a result, A2626 happens to be a potential candidate to study the interaction between the X-ray emitting ICM and the radio emitting plasma and also to assess the efficiency of the radio feedback from the AGN. This paper is organised as follows. Section~\ref{red} describes the observations and data preparation. In Section~\ref{sec3_result} we describe the results from the morphological and spectral analysis of the X-ray emission from this cluster. A discussion on the results from this analysis is presented in Section~\ref{discussion}, while brief summary of the study is given in Section~\ref{sec4_summary}.

In this paper, we assume a standard $\Lambda$ cold dark matter (CDM) cosmology, with H$_0$= 73 km s$^{- 1}$ Mpc$^{- 1}$, $\Omega_m$ = 0.27, and $\Omega_{\Lambda}$ = 0.73, so that at the redshift of 0.0553 \citep{1999ApJS..125...35S}, 1 arcsec corresponds to 1.01 kpc. Errors quoted in the spectral analysis are in the 68\% confidence limit and the metallicities are measured relative to those of \cite{1998SSRv...85..161G}. In all the images presented in this paper,  the north is to the top and east is to the left.\\

\begin{figure}
\centering
\includegraphics[scale=0.4]{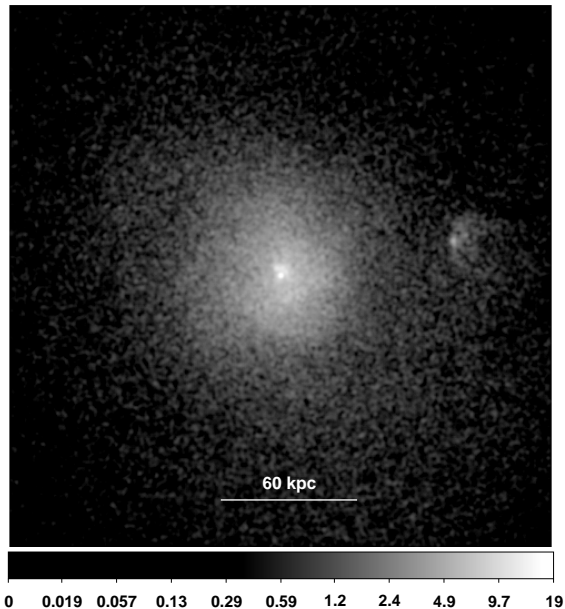}
\caption{Background subtracted, merged raw 0.5-3 keV \textit{Chandra} image of A2626. For better visualization this image has been smoothed with a 2$\sigma$ Gaussian kernel. The brighter source on the west of the central maximum is associated with a nearby member IC 5337. Isolated point sources were excluded for subsequent analysis.}
\label{fig1}
\end{figure}

\section[2]{Observations and Data Preparation}
\label{red}

A2626 was observed twice by \textit{Chandra} X-ray observatory on 2003 January 22 for 24.7 ks (ObsID 3192) and later on 2013 October 20 for 110.8 ks (ObsID 16136) in VFAINT mode with the source focused on the back illuminated ACIS-S3 chip (Table~\ref{obs}). Level 1 event files corresponding to both the observations were reprocessed using the \textit{chandra\_repro} task within \textit{CIAO V 4.9} using \textit{CALDB V 4.7.6} and were corrected for the latest charge transfer inefficiencies and time-dependent gain problems. The event files were also filtered for flaring events exceeding 20\% of the average background count rates using \textit{lc\_sigma\_clip} algorithm. System supplied background files appropriate to each of the observation were identified from the ``blank-sky" using \textit{acis\_bkgrnd\_lookup} tool and were normalized to match the 10-12 keV count rate in the science frames. Both these observations were merged after constructing their respective PSFs to produce a single clean event file of 134 ks. Finally, point sources detected in the merged image by the CIAO tool \textit{wavdetect} were removed and the holes due to their removal were filled in with the background emission using the tool \textit{dmfilth}. \\

\begin{figure}
\centering
\includegraphics[scale=0.4]{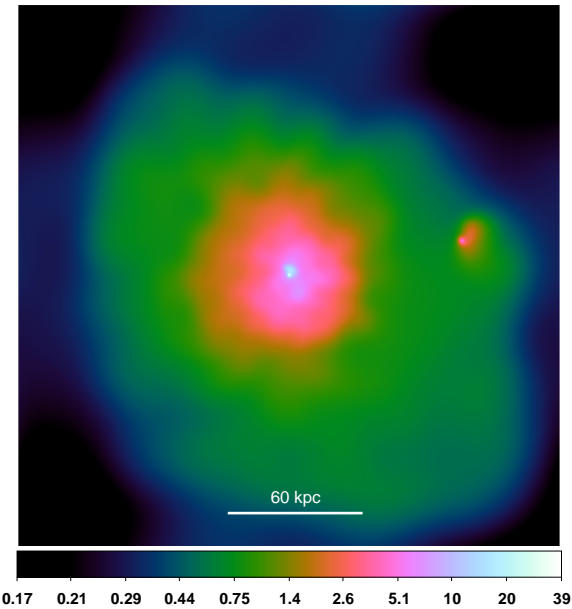}
\caption{0.5-3 keV adaptively smoothed \textit{Chandra} X-ray image of A2626. Notice the substructures evident in the central region of this cluster emission.}
\label{fig_3}
\end{figure}

\begin{figure}
\centering
\includegraphics[scale=0.4]{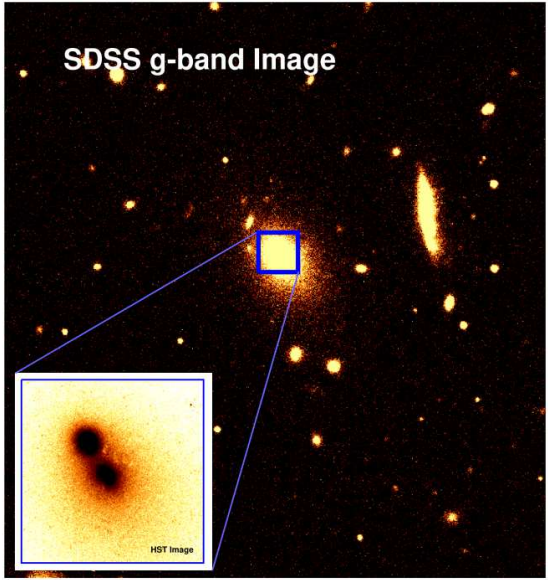}
\caption{The 5$\arcmin\times$5$\arcmin$ (60kpc$\times$60kpc) SDSS $g$-band image of A2626. Central 20$\arcsec\times$20$\arcsec$ of A2626 as observed by HST telescope is shown in the inset, which exhibits two nuclear sources associated with the CD galaxy.}
\label{fig2}
\end{figure}

The background subtracted, exposure-corrected, merged 0.5-3 keV \textit{Chandra} image of A2626 shown in Figure~\ref{fig1} reveals a number of features in the core region of the cluster and are consistent with those reported by other researchers \citep[e.g.,][]{2008ApJ...682..155W,2018A&A...610A..89I}. The brightest region near the centre is associated with the cD galaxy IC 5338, while the nearby galaxy IC 5337 is seen on its west at a separation of 85 kpc. Small fluctuations due to the presence of cavities, excess emission, edge, etc. are also apparent in the surface brightness distribution of the X-ray emission in Figure~\ref{fig_3}. For better visualization this image is smoothed with a 2$\sigma$ Gaussian kernel. Figure~\ref{fig2} represents 5$\arcmin\times$5$\arcmin$ SDSS $g$-band image of the central region of this cluster, where the cD galaxy IC 5338 and the near by member IC 5337 on its west are clearly visible. The inset shows central 20$\arcsec\times$20$\arcsec$\, region of A2626 imaged using HST delineating a pair of bright sources in the core of this cluster. The NE (north-east) source WHL J233630.6+210850 \citep[$z$ = 0.0546, RA: 23h36m30.6s, $\delta$: +21d08m50s;][]{2013MNRAS.436..275W} and the SW (south-west) source IC 5338 \citep[$z$ = 0.0546, RA: 23h36m30.4s, $\delta$:+21d08m46s;][]{2009A&A...495..707C} are seen at a separation of $\sim$ 3.4 kpc.\\ 

\section[3]{Results}
\label{sec3_result}
\subsection{X-ray imaging analysis}
\label{sec3.1_img}
\begin{figure*}
\centering
{
\includegraphics[scale=0.42]{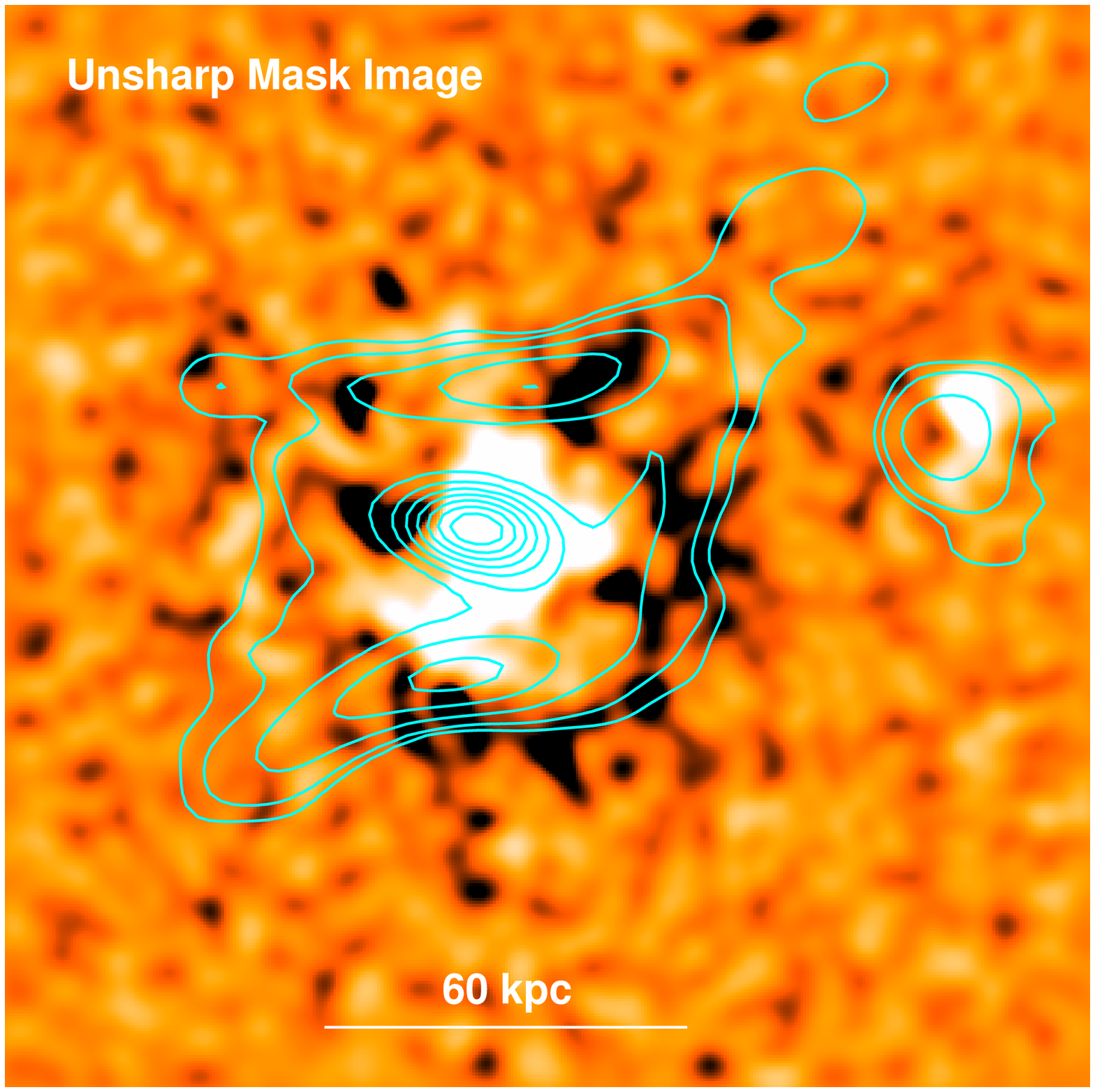}
\hspace{1cm}
\includegraphics[scale=0.53]{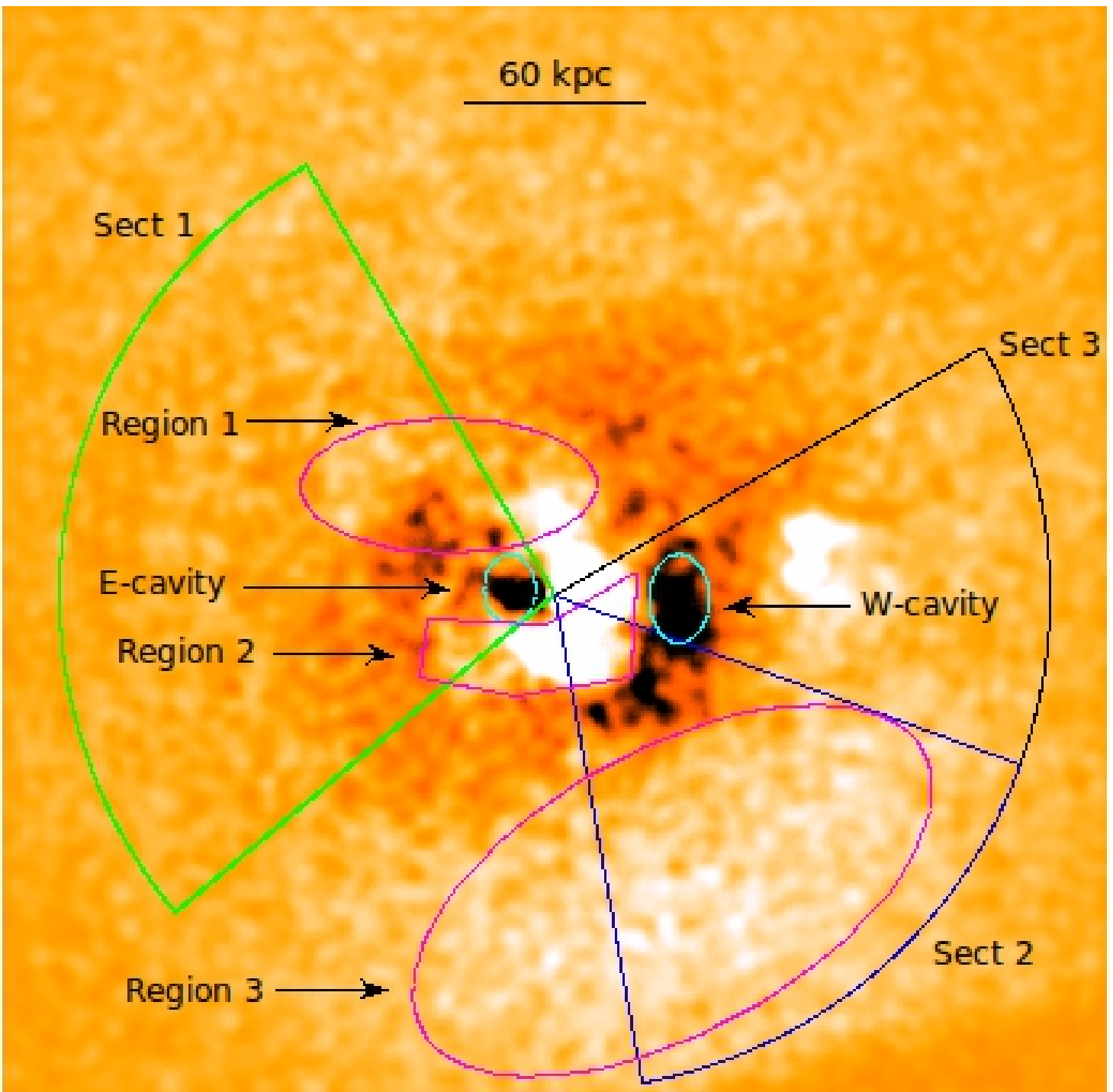}
}
\caption{{\it left panel}: 0.5-3 keV \chandra unsharp mask image of A2626 obtained after subtracting a 18$\sigma$ wide Gaussian kernel smoothed image from that smoothed with 2$\sigma$ Gaussian kernel. Overlaid on it are the 1.45 GHz VLA radio contours (cyan colour) to check the association of radio emission with the X-ray deficient regions (cavities). The 3$\sigma$ radio contours are at 0.13, 0.21, 0.48, 0.92, 1.53, 2.33, 3.29, 4.44, 5.75 and 7.25 mJy~beam$^{-1}$. The radio image rms noise is 43 $\mu$Jy~beam$^{-1}$ at the resolution of 9\arcsec\,$\times$ 9\arcsec. Notice the substructures in the X-ray surface brightness distribution. Darker shades represent the X-ray deficient regions while brighter are due to the excess emission. {\it right panel}: 0.5-3 keV 2D smooth model subtracted residual map revealing hidden substructures embedded within the surface brightness distribution in the environment of A2626. A pair of X-ray cavities (the E and W-cavity) is clearly seen. The regions of interests used for the spectral extraction are overlaid on the residual image. Purple ellipses and the polygon indicates the regions of excess emission. Three sectorial regions used for the surface brightness and spectral analysis are also highlighted in this figure.}
\label{fig3}
\end{figure*}

To investigate the presence of X-ray deficit bubbles or cavities and other hidden features within diffuse X-ray emission we construct 0.5-3 keV unsharp-mask image of the combined \chandra data. This was achieved by subtracting an image smoothed at wider Gaussian kernel from that smoothed with a narrow kernal. We tried several combinations of the Gaussian kernel widths to smooth the images for deriving the unsharp-mask, however, the one obtained after subtracting 18 arcsec wide Gaussian kernel smoothed image from that with 2 arcsec yielded the better visibility of the hidden features. The resultant unsharp-mask image (Figure~\ref{fig3}, {\it left panel}) clearly reveals a number of features that otherwise were not apparent in the cleaned X-ray image. The central excess emission (brighter shade) is seen elongated in the north-south direction, while a pair of depressions (darker shades) due to the presence of X-ray cavities are on the east (E-cavity) and the west (W-cavity) of the peak emission. The excess X-ray emission seen on the west of the central source is due to  a nearby member IC~5337. The 1.4 GHz radio contours from VLA data overlaid on this image exhibit association of the southern radio arc with the arc shaped X-ray deficit region in the unsharp-mask image and confirms the findings of \cite{2016ApJS..227...31S} and \cite{2018A&A...610A..89I}. \\

The features seen in the unsharp-mask were further confirmed in the residual map (Figure~\ref{fig3} \textit{right panel}) obtained after subtracting 2D smooth model of the X-ray emission in the cleaned \textit{Chandra} image of A2626. The smooth model was obtained by fitting a 2D double $\beta$-model to the surface brightness distribution in the cleaned 0.5-3 keV X-ray image. This image clearly exhibit two X-ray deficit regions, the E-cavity and the W-cavity, at $\sim$13 and 39 kpc, respectively. Presence of the E-cavity was already reported by \cite{2008ApJ...682..155W} and \cite{2016ApJS..227...31S}, however, the recent deep observations have exhibited presence of the W-cavity in the surface brightness distribution. \citet{2018A&A...610A..89I} could also detect the E-cavity, however, they suspect it presence due to the artifact of the asymmetric X-ray emission. We also confirm their presence through the X-ray count analysis. For this we have extracted 0.5-3 keV counts from the arc shaped annuli  covering these cavity regions and dividing them into several sectors. Different regions of interests including X-ray cavities and excess emission are highlighted in this figure. Like the findings of \cite{2008ApJ...682..155W} this image reveals an arc shaped excess emission on the north and the south of X-ray peak (Region 1 and 2) of the cluster. We also find the extended excess X-ray emission on the south-west (Region 3) of the X-ray peak as reported by \cite{2008ApJ...682..155W}. 

To examine overall morphology of the hot gas distribution within A2626 we also derive its azimuthally averaged surface brightness profile of the X-ray emission. For this we extract the mean surface brightness in 2 kpc wide concentric circular annuli centred on the X-ray peak of the exposure corrected, background subtracted 0.5-3 keV X-ray image of A2626. The resulting surface brightness was then fitted with the standard $\beta$-model \citep{1976A&A....49..137C}
\begin{equation}
S(r) = S_0 \left[ 1 + \left(\frac{r}{r_c} \right)^2 \right]^{- 3 \beta + 0.5}
\end{equation}

here, $S_0$ and $S(r)$, respectively, represent the central surface brightness and at the projected distance $r$, $\beta$ the slope parameter depending on the ratio of the specific energy content of the dark matter and of the gas and $r_c$ represents the core radius. This model assumes that the X-ray emitting ICM and the galaxies within a cluster are in hydrostatic equilibrium and follow isothermal distribution. The single $\beta$-model fit yielded $\beta$ = 0.4 and $r_c$=13.6$\pm$0.2 kpc, and is in agreement with that obtained by \citet{2018A&A...610A..89I}. However, deviates largely from that obtained by \citet{2008ApJ...682..155W} ($\beta\sim$0.7). Usually, the standard model expects this parameter to take value close to $\beta\sim\,2/3$ while constraining the ICM distribution in outer parts of relaxed clusters. As A2626 is a cool core cluster exhibiting excess X-ray emission above that expected by the standard model in the central region, therefore, single $\beta$-model fails to give appreciable fit. For the cooling flow clusters, where non-gravitational heating is operative, this excess emission drives the core radius and $\beta$ value to taken low values \citep{2000MNRAS.315..356H}. As a result the slope parameter for the case of A2626 has deviated by a large factor. There are several cases of cool core clusters for which the reported slope parameter values are as low as 0.37 \citep[e.g.,][]{1999ApJ...520...78H,2006PASJ...58..131F,2013MNRAS.428...58R}, and are related to the cooling flows \citep{1993Natur.363...51P,1994ApJ...428..544D}. The larger value obtained by \citet{2008ApJ...682..155W} is due to the reason that they continue this fit out upto $\sim$800 kpc. Larger the fitting radius, the $\beta$-parameter takes larger value \citep[close to unity;][]{1996MNRAS.283..431B}.

\begin{figure}
\centering
\includegraphics[scale=0.45]{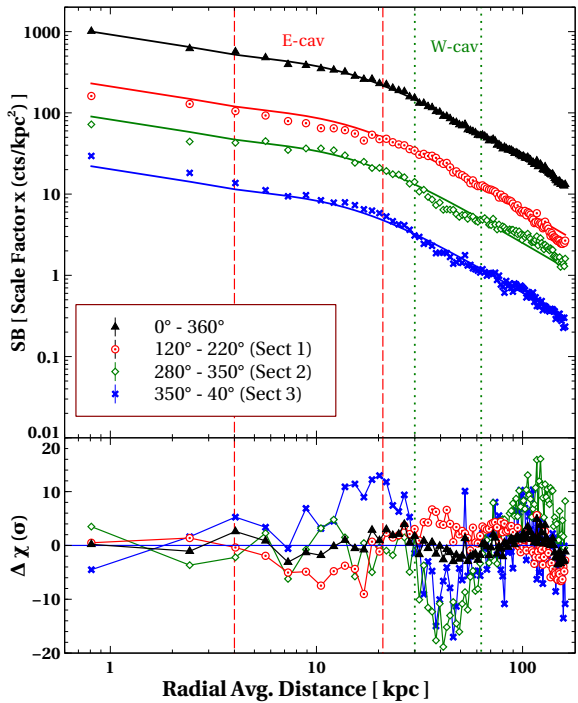}
\caption{Azimuthally averaged X-ray surface brightness (black triangles) fitted with a double $\beta$ model (continuous line) for the gas emission from A2626. Surface brightness profiles extracted along three wedge shaped sectors covering 120$^o$-220$^o$(red circles), 280$^o$-350$^o$ (green diamonds) and 350$^o$-40$^o$ (blue crosses) are also shown in this figure. For better visualization the profiles are shifted arbitrarily along Y-axis. Lower panel depicts residuals between the data and the model.}
\label{fig4}
\end{figure}

As the single $\beta$-model yielded larger residuals in the inner region, to constrain the emission in a proper way, we added one more $\beta$ component for the fitting purpose \citep{2000MNRAS.318..715X}
\begin{equation}
S (r) = S_0{_1}  \left[ 1 + \left( \frac{r}{r_c{_1}} \right)^2 \right]^{- 3 \beta{_1} + 0.5} + S_0{_2} \left[ 1 + \left( \frac{r}{r_c{_2}} \right)^2 \right]^{- 3 \beta{_2} + 0.5}
\end{equation}
The slope parameters and the core radii for the resultant fit were then found to be 0.4, 0.7; and 2.2$\pm$1.1 kpc, 13.6$\pm$0.2 kpc, respectively, which are in good agreement with the standard model. The resultant azimuthally averaged surface brightness profile (black triangles) along with the best fit double beta model (the continuous line) are shown in Figure~\ref{fig4}.

Though the double beta model has constrained the cluster emission appreciably, small derivations seen at some of the radii are probably due to the presence of X-ray cavities and excess emission. To investigate these features in further detail we compute three more surface brightness profiles by extracting X-ray emission from three different conical sectors (Figure~\ref{fig3} \textit{right panel}) namely, Sect1 (covering the E-cavity i.e., 120$^o$-220$^o$), Sect2 (covering the excess emission Region 3, 280$^o$-350$^o$) and Sect3 (the W-cavity,  350$^o$-40$^o$) and are shown in the same figure. For comparison we also plot the azimuthally averaged best-fit model on top of sectorial surface brightness profiles and for better visualization are shifted along the ordinate axis by arbitrary values. The depressions in the surface brightness profiles along Sect1 and Sect3 at 13.1 kpc and 38.9 kpc, respectively, are due to the X-ray cavities, while the edges at $\sim$36 kpc and 33 kpc along Sect2 and Sect3, respectively, corresponds to the cold fronts (discussed in Section~\ref{sec4.5_proffit}).

\subsection{X-ray spectral analysis}

\begin{figure*}
\centering
{
\includegraphics[scale=0.55]{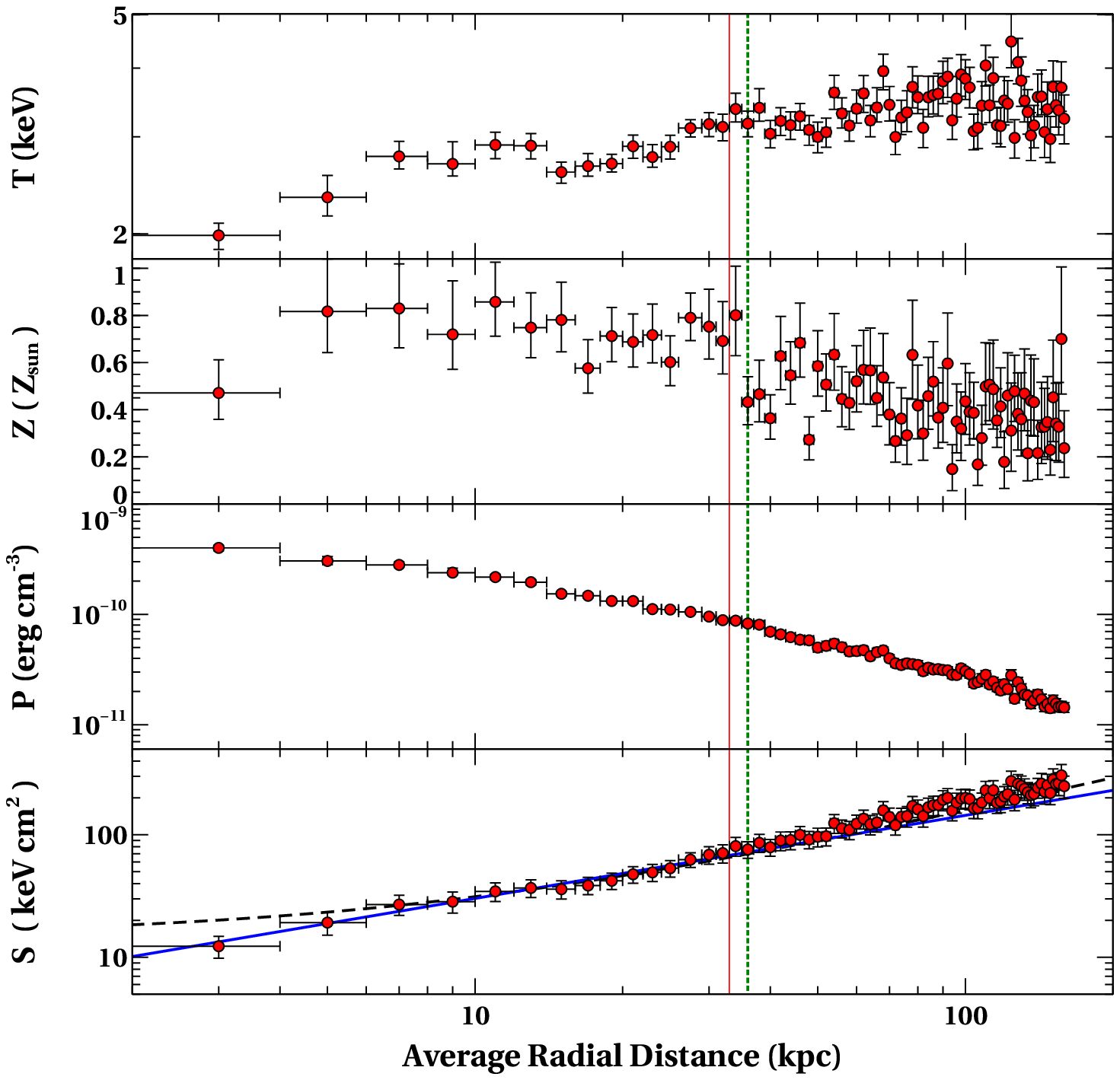}
\includegraphics[scale=0.55]{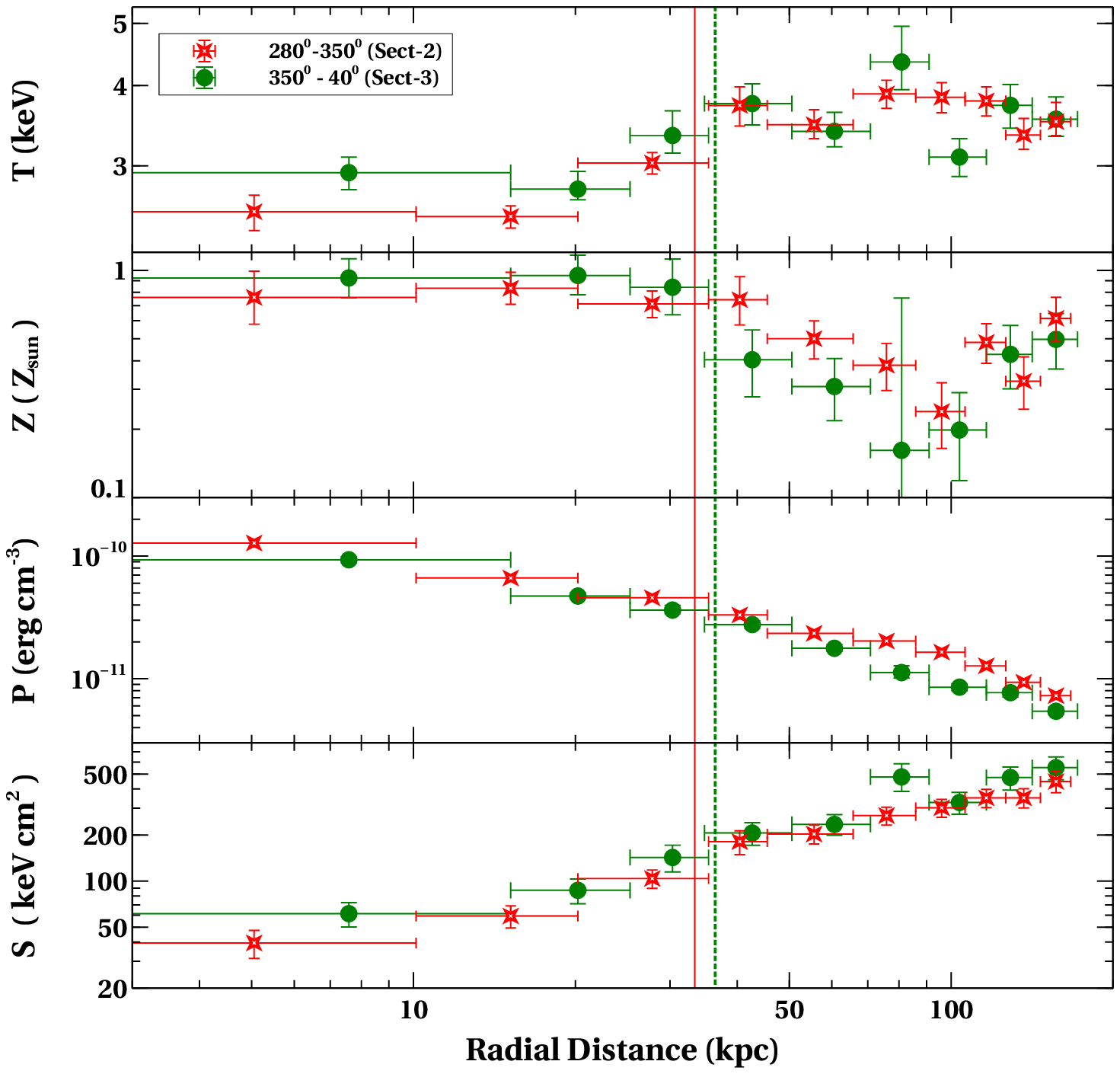}
}
\caption{\textit{left panel:} Projected azimuthally averaged radial temperature profile of the gas within A2626 (first row). Like other cooling flow clusters, temperature within A2626 exhibits an increasing trend as a function of the radial distance.  We also plot the best fit metallicity (row 2), pressure (row 3) and the entropy (row 4) in the same figure. The blue continuous line in the entropy profile represents $S(r)=S_0 + S_{100}(r/100 kpc)^{\alpha1}$, while the grey line represent $S(r)=S_{100}(r/100 kpc)^{\alpha2}$. Notice the floor or excess in the entropy in the core region relative to the lowest value (ideally zero). \textit{right panel:} Same as that in the left panel.  The red data points are for the extraction from the sectorial region with opening angles 280\degr-350\degr (Sect 2), while the green points are for the 350\degr-40\degr (Sect 3). Notice the drop in the temperature just beyond 30 kpc in both the sectors and are probably associated with the edge in the surface brightness along them. The vertical line indicate the location of the cold fronts.}
\label{azmuth_spec}
\end{figure*} 

\begin{figure*}
\centering
{
\includegraphics[scale=0.4]{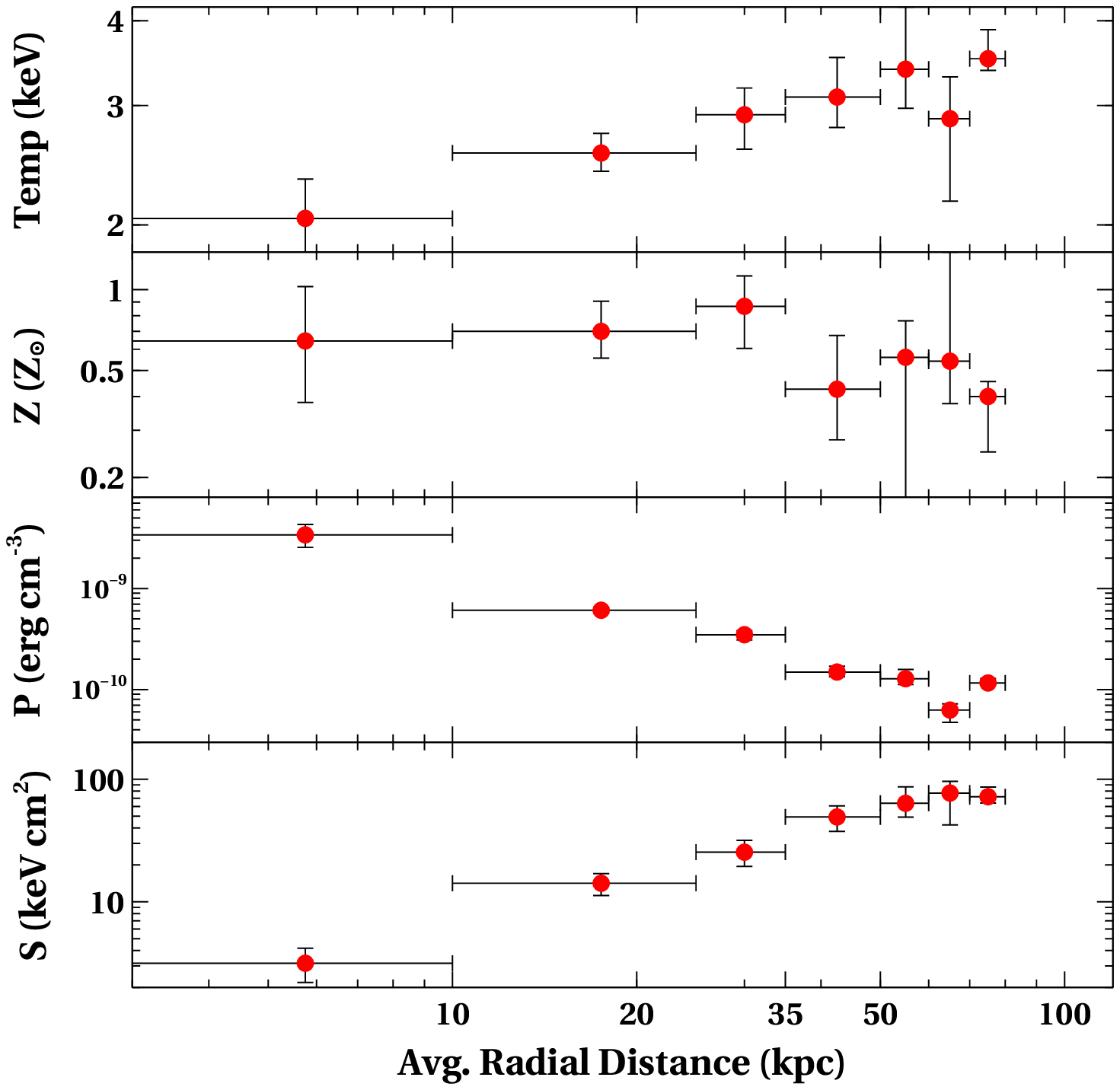}
}
\caption{Azimuthally averaged, deprojected profiles of the temperature and other thermodynamical parameters. This method follows ``onion peeling" analogy to remove contribution from the overlying layers of the ICM \citep{2003ApJ...585..227B}.}
\label{fig_depro}
\end{figure*}


To investigate the radial dependence of the ICM properties, we compute azimuthally averaged profiles of the thermodynamical parameters by extracting 0.5-7 keV spectra from 78 different concentric annuli of widths 2 kpc, centred on the X-ray peak of the cluster emission. The annuli were constructed from ObsID 16136 and yielded $\sim$4000 counts, except in the central region where the extracted counts were $\sim$1000. Spectra were also extracted from ObsID 3192 using the same annuli as obtained for ObsId 16136. For background treatment we used the system supplied blank-sky frames. Spectra extracted independently from each of the observation were then fitted simultaneously with a single temperature collisional equilibrium plasma model {\sc apec} \citep{2001ApJ...556L..91S} modified by the foreground absorption model {\sc wabs} within {\sc XSPEC} V 12.9.1, with the foreground absorption column density fixed at $N_H = {\rm 4.3 \times 10^{20}\,cm^{-2}}$ \citep{2005yCat.8076....0K}.

We then compute the electron density $n_e$ (in units of cm$^{-3}$) from the {\sc apec} normalization as \citep{2001ApJ...556L..91S}
\begin{equation}
n_e = \sqrt{\frac{1.2\times 10^{14}(4\pi D_A^2) \times {\rm norm}}{(1+z)^2 V}}
\end{equation}
where, $D_A$ and $V$ represent the angular diameter distance of the source and volume of the spherical shell, respectively. For the fully ionized gas with hydrogen and helium mass fractions of X = 0.7 and Y = 0.28, we assume $n_e\sim 1.18 n_H$. We also calculate the pressure and the entropy of the gas within individual annulus as $p=nkT$ and $S=kTn_e^{-2/3}$, respectively, assuming an ideal gas density of $n\sim 1.92 n_e$. 

The resulting profiles of the thermodynamical parameters temperature, metallicity, pressure and entropy of the ICM are shown in left panel of Figure~\ref{azmuth_spec}. A number of features are apparent in these profiles. The temperature profile exhibits a rise in the ICM temperature in the outer direction and hence confirms the typical characteristic of a cooling flow in the core of A2626 \citep{2007ApJ...669..158G,2009ApJ...705..624D,2009ApJ...693.1142S,2012MNRAS.421.1360H,2015Ap&SS.359...61S,2016MNRAS.461.1885V,2017MNRAS.466.2054V,2018MNRAS.474.1065S}. Apart from this systematic rise, the temperature profile also exhibit discontinuity at $\sim$33 kpc, and is probably linked with the edge in the surface brightness (Figure~\ref{fig4}). \citet{2018A&A...610A..89I} have also reported an inward jump in temperature at this location. The metallicity profile shows a dip in the innermost bin, taking value lower than one-half of the solar abundance, which then rises to about 0.9 solar value and then drops to one-half of solar value in the outskirt. A small drop in it is seen at $\sim$33 kpc.

\begin{table*}
{\footnotesize  {\footnotesize {\hspace{2mm}} 
\caption{Thermodynamical parameters of the X-ray emission from different regions of interests within the central region of A2626}.
\begin{tabular}{|c|c|c|c|c|c|}
  \hline
  Regions		&  Region 1  &  Region 2  &
  Region 3   & East cavity  & West cavity \\
  \hline
  Temperature (keV)  & 2.95$\pm$0.07  & 2.89$^{+ 0.05}_{- 0.06}$  &
  3.48$\pm$0.08  & 3.17$^{+ 0.13}_{- 0.13}$  & 3.62$^{+ 0.27}_{- 0.27}$ \\
  
  Abundance (\Zsun)  & 0.46$\pm$0.04  & 0.70$\pm$0.04  & 0.32$\pm$0.03  &
  0.62$^{+ 0.11}_{- 0.10}$  & 0.47$^{+ 0.13}_{- 0.13}$ \\
  
  Pressure (10$^{- 9}$ erg cm$^{-3}$)  & 2.02$\pm$8.2  & 2.03$\pm$7.8  & 3.19$^{+ 1.3}_{-
  1.3}$  & --  &  -- \\
  
  Electron density (cm$^{-3}$)  & 0.21$\pm$0.03  & 0.22$\pm$0.03  & 0.29$\pm$0.03  &
  --  & -- \\

  Entropy (keV cm$^{2}$)  & 8.18$\pm$8.3  &  7.85$\pm$0.7  & 7.95$^{+ 0.66}_{- 0.67}$  &
  -- & -- \\
  \hline
\end{tabular}\label{spec_tab}}}
\end{table*}

The radial nature of the gas entropy in this cluster environment (row 4 of Figure~\ref{azmuth_spec}) was then fitted with an empirical model $S(r) = S_{100}(r/100 kpc)^{\alpha}$ (the grey line). Here, $S_{100}$ represents the gas entropy at 100 kpc and the exponent factor $\alpha$. The model could fit the data appreciably in the outer region, however, showed deviation in the central part of the cluster. To constrain it we fitted the data with $S(r)=S_0 + S_{100}(r/100 kpc)^{\alpha1}$ (shown by the blue continuous line), where $S_0$ represents an additional component of the entropy at the core region, which modeled the data appreciably. This plot clearly imply that the entropy in the central region of the cluster is not reaching to the lowest value (ideally zero), instead exhibits a floor at $\sim$14.8 keV cm$^2$ and towards some intermittent heating imply at the core. 
As the surface brightness profiles along Sect2 and Sect3 show edges, therefore, to confirm their association with the shocks and cold fronts we also derive profiles of the thermodynamical parameters of the X-ray emission from these two sectors. For this we extract 0.5-7 keV X-ray photons from different sectorial bins of width $\sim$10 kpc from the wedge shaped sectors with opening angles 280$^o$-350$^o$ (Sect2) and 350$^o$-40$^o$ (Sect3). The extracted spectra were then treated in the same way as above and profiles of the resultant parameters are shown in right panel of Figure~\ref{azmuth_spec}. The temperature profiles along these sectors revealed small jumps at 36 kpc along Sect2 and at 33 kpc along Sect3. We also derive profiles of metal abundance, pressure ($p$) and entropy ($S$). Metallicity profile along the Sect2 shows a drop in the bin next to 30 kpc, while the pressure and the entropy profiles showed discontinuity at this location. 

We also derive the azimuthally averaged, deprojected  profiles of the temperature and other thermodynamical parameters. The deprojection analysis assumes the contribution from the projected three dimensional spherical shells into the two-dimensional circular annuli and removes them following the ``onion peeling'' analogy of \cite{2003ApJ...585..227B}. Here, the X-ray counts from the overlying layers are removed from the successive shells and hence yields very few counts and therefore required us to increase the bin size to reach the required statistics. This method is used for the spectral investigation because the fitting is affected by the projection effect. Like above, spectra extracted from the individual annulus were fitted simultaneously using the models {\sc wabs}*{\sc apec} and the results are shown in  Figure~\ref{fig_depro}. A small discontinuity is apparent at $\sim$33 kpc in the temperature and the metallicity profiles. The past studies report several clusters hosting cold fronts with an associated discontinuity in the metallicity profile and an absence of clear discontinuity in the temperature profile (e.g., A2052: \citealp{2010A&A...523A..81D}; A2199: \citealp{2006MNRAS.371L..65S}; 2A 0335+096: \citealp{2003ApJ...596..190M}). The larger errorbars in the resultant profiles are due to fewer counts in the deprojected shells.

We also perform the  spectral analysis of the 0.5-7 keV  X-ray photons from different regions of interests shown in  Figure~\ref{fig3} (\textit{right panel}) and were fitted independently using the same model. The resultant thermodynamical parameters are given in Table~\ref{spec_tab}. Here, X-ray spectra extracted from the ellipsoidal shaped X-ray cavities, discussed in Section~\ref{sec4.1_cav} revealed that the plasma that fills the cavities is relatively hotter than its surroundings. X-ray emission from Region 3 is also relatively hotter than that from other excess regions.

\section[4]{Discussion}
\label{discussion}
\subsection{Cavity energetics}
\label{sec4.1_cav}
X-ray bubbles or cavities in the unsharp mask and residual map of A2626 are likely formed due to the interaction between the radio jets launched by the central AGN and the surrounding ICM. During an AGN outburst, the radio jets do $p$V work on the ICM and inflate cavities, which then rise buoyantly in the wake of the ICM until they reach pressure equilibrium. Once they reach the equilibrium, where buoyant velocity exceeds the expansion velocity, get detached from the jets and hence can supply their enthalpy to the ICM \citep{2004ApJ...607..800B}. Thus, the systematic studies of X-ray cavities in cool core clusters act as a calorimeter to quantify the mechanical power that has been delievered by the central AGN into the ICM without requiring its radio observations. \textit{Chandra} due to its high angular resolution has provided us with a most reliable way of measuring the jet power in several galaxy clusters through the cavity analysis. The total enthalpy i.e. sum of the $p$V work and the internal energy that provides required pressure to support the cavities was quantified following the standard procedure outlined by \cite{2004ApJ...607..800B} 
\begin{equation}
  E_{cav} = \left[ \frac{\gamma}{\gamma - 1} \right] pV
\end{equation}
where, $p$ is the thermal pressure of the ICM at the cavity position and was obtained from the projected radial profile derived above, V the volume of the cavity and $\gamma$ the adiabatic index of the fluid filling the cavity. Assuming that the cavities are filled with relativistic plasma i.e., $\gamma$=4/3, we estimate the enthalpy content of each of the cavity as E$_{cav}$ = 4$p$V. 

Assuming the X-ray bubbles as ellipsoids of the projected semi-major and semi-minor axes $R_l$ and $R_w$, respectively, oriented along the jet we estimate volume of each of the cavity as $V=4\pi {R_w}^2 R_l/3$. As depths of the bubbles are not known, therefore, we assume cavities as prolate ellipsoids for estimating their volumes. The measured values of the cavity sizes, volumes, and per cavity enthalpy content (4$p$V)  are given in Table~\ref{cav_energy}. The accuracy in the measurement of the average enthalpy content of  cavities strongly depend on the uncertainties in the volume measurement. In the present study the cavity dimensions ($R_l$, $R_w$) and hence volumes were measured by the visual inspection of the unsharp mask image, therefore, are prone to systematic errors. They are also strong functions of the data quality and are affected by the projection effect in determining spatial geometry of the cavity. As a result, volumes measured by different researchers may vary significantly and hence may lead to large uncertainties in the measurement of the total enthalpy content of the cavities \citep{2010ApJ...714..758G}. In the present study we assume the errors in the volume measurement to be $\sim$20\%.

\begin{table}
\centering
\scriptsize
\caption{Details of the energy contents of the X-ray cavities in the environment of A2626}
\footnotesize{
\hspace{2mm}
\begin{tabular}{| c |c |c |c |c |c |c |c|}
\hline
{Cavity Parameters} & East cavity & West cavity   \\
\hline	
R$_1$ (kpc)               &11.34               &14.87                 \\

R$_w$ (kpc)               &8.5               &9.91                  \\

R (kpc)          &13.07      &38.95                  \\

Cavity vol (cm$^3$)        &1.17$\times$10$^{69}$ &2.36$\times$10$^{69}$  \\
 

 

t$_{age}$ (yr)          & 1.7$\times$10$^7$         &8.4$\times$10$^7$        \\

4$pV$ (erg)              & 2.10$\times$10$^{59}$          &2.5$\times$10$^{59}$     \\

P$_{cavity}$ (erg s$^{-1}$)    & 9.66$\times$10$^{44}$  & 3.61$\times$10$^{44}$      \\
\hline 
\end{tabular}
}
\label{cav_energy}
\end{table}

We then estimate the rate at which AGN injects its energy into the cluster environment as $P_{cav} = E_{cav} / t_{age}$, where, age of the cavity $t_{age}$ was estimated assuming the buoyancy rise time of the bubbles \citep{2004ApJ...607..800B}. The cavity age is defined as the time taken by the cavity to reach its terminal velocity and is given by 
\begin{equation}
t_{age} \approx t_{\rm buoyancy} \approx R \sqrt{\frac{AC_D}{2gV}}
\end{equation}
Here, $R$ represents the projected distance of the cavity from the cluster centre, $S$ the cross sectional area of the cavity ($S=\pi {R_w}^2$), and $C_D$ the drag coefficient \citep[$\sim$0.75 in the present case,][]{2001ApJ...554..261C}. This analysis yielded the average cavity power to be equal to ${\rm P_{cav} \sim 6.6 \times 10^{44}\,erg\,s^{-1}}$, and was utilized for heating up of the ICM. 

\subsection{Heating vs cooling of the ICM}
As was mentioned earlier, one of the primary goals of this paper was to check whether the power supplied by the AGN outburst is sufficient enough to balance the radiative loss of the ICM. To establish this we derive the cooling time profile using above derived azimuthally averaged profiles of temperature, metallicity and electron density. The cooling time $t_{cool}$ of a gas parcel with electron density $n_e$, temperature $T$ and metallicity $Z$ is defined as the time required for the gas to radiate out its enthalpy and is given by \cite{1986RvMP...58....1S},

\begin{equation}
  t_{cool} = \frac{5}{2}\frac{kT}{\mu X_H n_e \Lambda(T)} = 8.5 \times 10^{10}
  \left( \frac{n_e}{10^{- 3} cm^{- 3}} \right)^{- 1} \left( \frac{T}{10^8
  \text{\textrm{K}}} \right)^{0.5}
\end{equation}

\begin{figure}
\centering
\includegraphics[scale=0.5]{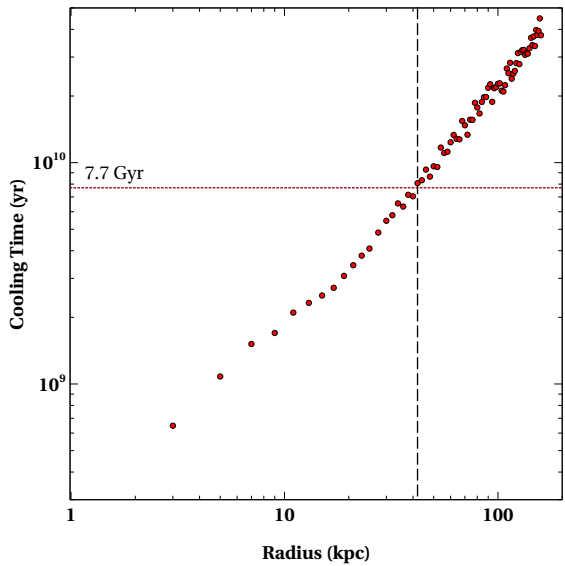}
\caption{Cooling time profile of the ICM in A2626. Horizontal dotted line indicate the cosmological age of the cluster 7.7 Gyr}
\label{cooling_time}
\end{figure}

where $\mu$ = 0.61 represents the molecular weight for a fully ionized plasma, $X_H$ = 0.71 the hydrogen mass function, and $\Lambda(T)$ the cooling function. The resultant cooling time profile of the ICM shown in Figure~\ref{cooling_time} implies that the ICM in the core region ({\rm $\sim14$} kpc) must have cooled down to the molecular phase a long ago. Here, the cooling radius ($r_{cool}$) is defined as the radius at which $t_{cool}$ becomes equal to the age of the system, the look-back time at $z$=1 and is taken as $\sim$ 7.7 Gyr \citep{2012MNRAS.427.3468B}. The present analysis yielded the cooling radius equal to $\sim$ 42 kpc and the X-ray cooling luminosity equal to ${\rm L_{cool} = 2.30\pm0.02 \times 10^{43}\,erg\,s^{-1}}$. 

\begin{figure*}
\centering
{
\includegraphics[scale=0.8]{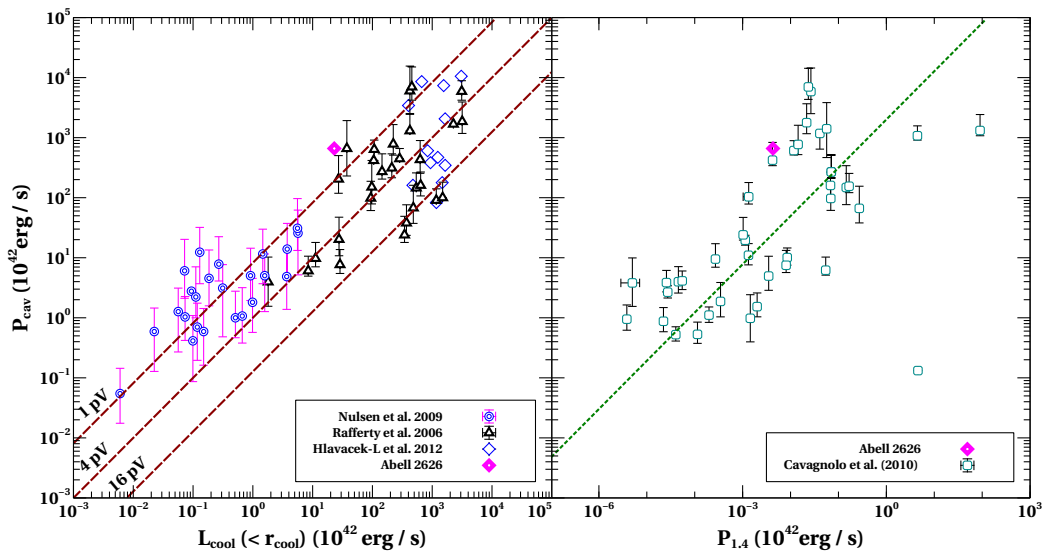} 
}
\caption{{\it left panel:} Balance between the cavity power vs. X-ray luminosity within the cooling radius obtained from \citet{2006ApJ...652..216R} (open triangles). For comparison, position of A2626 is plotted in the same plot and is denoted by the magenta diamond. The diagonal lines indicate the equivalence between $P_{cav}\,\&\,L_{cool}$ for the enthalpy levels $pV$, 4$pV$ and 16$pV$. For comparison, we also plot the positions of other cool core systems from \citet{2009AIPC.1201..198N, 2012MNRAS.421.1360H,2013MNRAS.431.1638H}. {\it right panel:} Balance between the cavity power (P$_{cav}$) vs 1.4\,GHz radio power for the sample studied by \citet{2010ApJ...720.1066C}. The dashed line represents the best fit relation for their sample of giant ellipticals (gEs). A2626 occupies position well above the best-fit line by \citet{2010ApJ...720.1066C}.}
\label{cav_re}
\end{figure*}

To establish whether the power injected by AGN outburst through the radio jets is sufficient enough to stop the cooling of the ICM in this cluster we check the balance between the mechanical power measured from the cavity analysis (P$_{cav}$) with the radiative loss of the ICM from within the cooling radius (${\rm L_{cool}(<r_{cool}}$)). For this we use Figure~6 of \cite{2006ApJ...652..216R}, where they plot the cavity (mechanical) power against the total radiative luminosity of the ICM (Figure~\ref{cav_re} \textit{left panel}). The diagonal lines  in this figure, respectively from top to bottom, represent the equality between them at the heat input rates of 1$p$V, 4$p$V and 16$p$V. A2626 in this figure occupies position above the 1$p$V line, implying that the `bubbling mode' of the  AGN outburst deposits about $\sim$29 times more power than required to offset the cooling flow in its core. In this figure for comparison we also show positions of other cool core clusters reported by other researchers \citep[e.g.,][]{2006ApJ...652..216R,2009AIPC.1201..198N,2012MNRAS.421.1360H,2013MNRAS.431.1638H}. 

Another convincing evidence for the confirmation of the balance between the heating versus cooling of the ICM was provided by the correlation between the mechanical power measured from X-ray cavity analysis (P$_{cav}$) and the 1.4 GHz radio power of the central source \citep{2008ApJ...686..859B,2010ApJ...720.1066C}. As these cavities are being created by the central AGN via radio jets and are being filled with relativistic plasma, therefore one expects an obvious correlation between the radio power and the cavity power. Further, as synchrotron losses at low frequency 327 MHz are very low relative to that at high frequency 1.4 GHz, therefore, a strong correlation is expected between the AGN power and the cavity power. However, 327 MHz is more sensitive to the extended emission rather than the jet power alone, while 1.4 GHz traces the minimum jet power. Therefore, for the case of AGN lifted X-ray bubbles \citet{2008ApJ...686..859B} observed a better correlation between the 1.4 GHz radio power and the mechanical power estimated through X-ray cavity analysis. As a result to check the balance between the heating versus cooling of ICM this correlation is a better probe. Similar correlation was also  reported by \citet{2016MNRAS.456.1172G} for the case of FR-I galaxies. They also find a weak correlation between the jet power and the radio luminosity. However, they warn that such correlations are strongly biased by the mutual distance dependence, therefore, must be taken with great care \citep[for details see][]{2016MNRAS.456.1172G}. With an objective to check the balance between the two measurements for A2626 we make use of the 1.4 GHz radio power versus cavity power correlation of \citet{2008ApJ...686..859B} and \cite{2010ApJ...720.1066C}. The best fit relation is shown by the dotted diagonal line. The 1.4 GHz monochromatic radio power ${\rm P_{1.4 GHz} \sim 4.30 \times 10^{39}\,erg\,s^{-1}}$ \citep{2004A&A...417....1G} and above measured mechanical power yielded A2626 to occupy position above the best fit relation in Figure~\ref{cav_re} (\textit{right panel}), implying that the mechanical power  estimated using cavity analysis is in excess of the 1.4 GHz radio power. However, it must be noted that the estimation of mechanical power include large ($\sim$ 20\%) uncertainties due to the inaccurate volume measurement, therefore, must be taken as the upper limit. 

\subsection{Radio emission} 
The wealth of the high resolution X-ray and radio data on a large sample of galaxy clusters has established that nearly all the  clusters with strong cool cores harbour a radio source at their core \citep{2006MNRAS.373..959D}.  A2626 is also reported to host a radio-mini halo with the 1.4 GHz diffuse radio emission spread in an atypical kite-like morphology with atleast three bright radio arcs along the north, west and south directions \citep{2005xrrc.procE8.12G,2013MNRAS.436L..84G}. \citet{2008ApJ...682..155W} predicted this unusual morphology of the radio emission likely due to the jet precession triggered by the gravitational interactions between the two sources associated with the cD galaxy. However, the later discoveries of the third arc on the west by \cite{2013MNRAS.436L..84G}, and the fourth arc on the east by \citet{2017MNRAS.466L..19K} using the low frequency 610 MHz GMRT data ruled out this model. \cite{2000A&A...356..788C} described this unusual morphology as result of the propagating buoyant radio emitting plasma that has been ejected by the central AGN in the form of subsonic plumes during its past activity.  

With an objective to examine the spatial correspondence between the X-ray bubbles and the  radio emission we overlay 1.4 GHz VLA radio emission contours on the X-ray unsharp mask image of A2626 (Figure~\ref{fig3} \textit{left panel}). The radio contours clearly reveal the atypical diamond shaped emission morphology even at low resolution. The central radio bar-like structure in the 1.4 GHz B configuration emission map presented by \cite{2004A&A...417....1G} roughly covers both the bubbles. Such a spatial connection between the E-cavity and the radio emission have also been detected by \cite{2018A&A...610A..89I}, however, they were cautious about claiming the cavity detection due to the reason that these are artifacts produced by the asymmetric X-ray emission. But the study by \cite{2016ApJS..227...31S} claims that the depression in the X-ray emission map is due to a cavity. \citet{2008ApJ...682..155W} also report similar association between the radio emission and the X-ray cavity, however, they find only E-cavity in the low S/N 24 ks \textit{Chandra} data (ObsID 3192). This analysis using 134 ks \chandra data on the basis of their appearance various image processing techniques, spatial association with the radio emission, balance between the cavity power and the observed radio power and cooling luminosity claim that these detection are genuine and carved by the radio jets launched by the central AGN. 

\subsection{Temperature and metallicity maps}

\begin{figure*}
\centering
{
\includegraphics[scale=0.35]{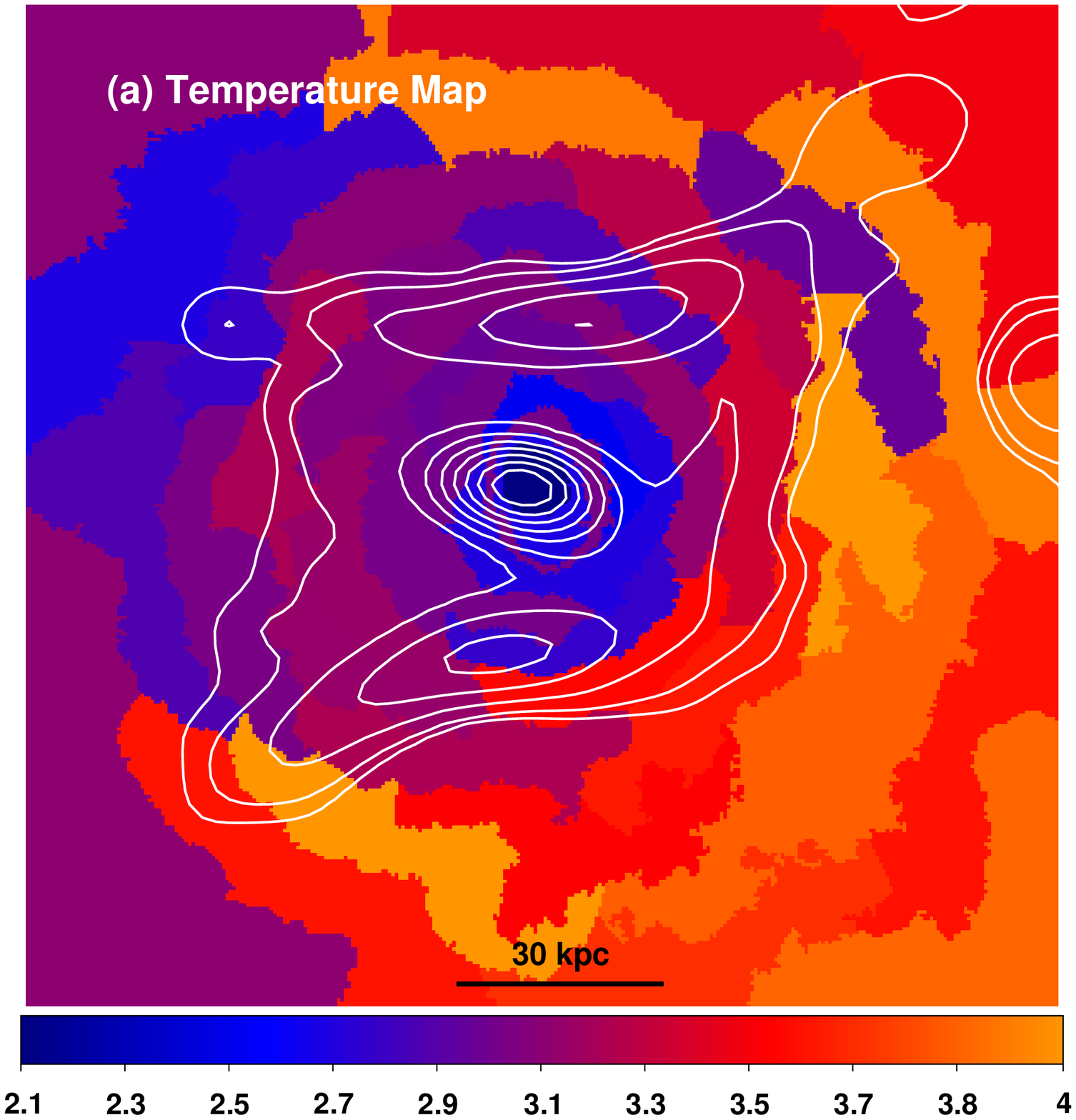}
\hspace{1cm}
\includegraphics[scale=0.35]{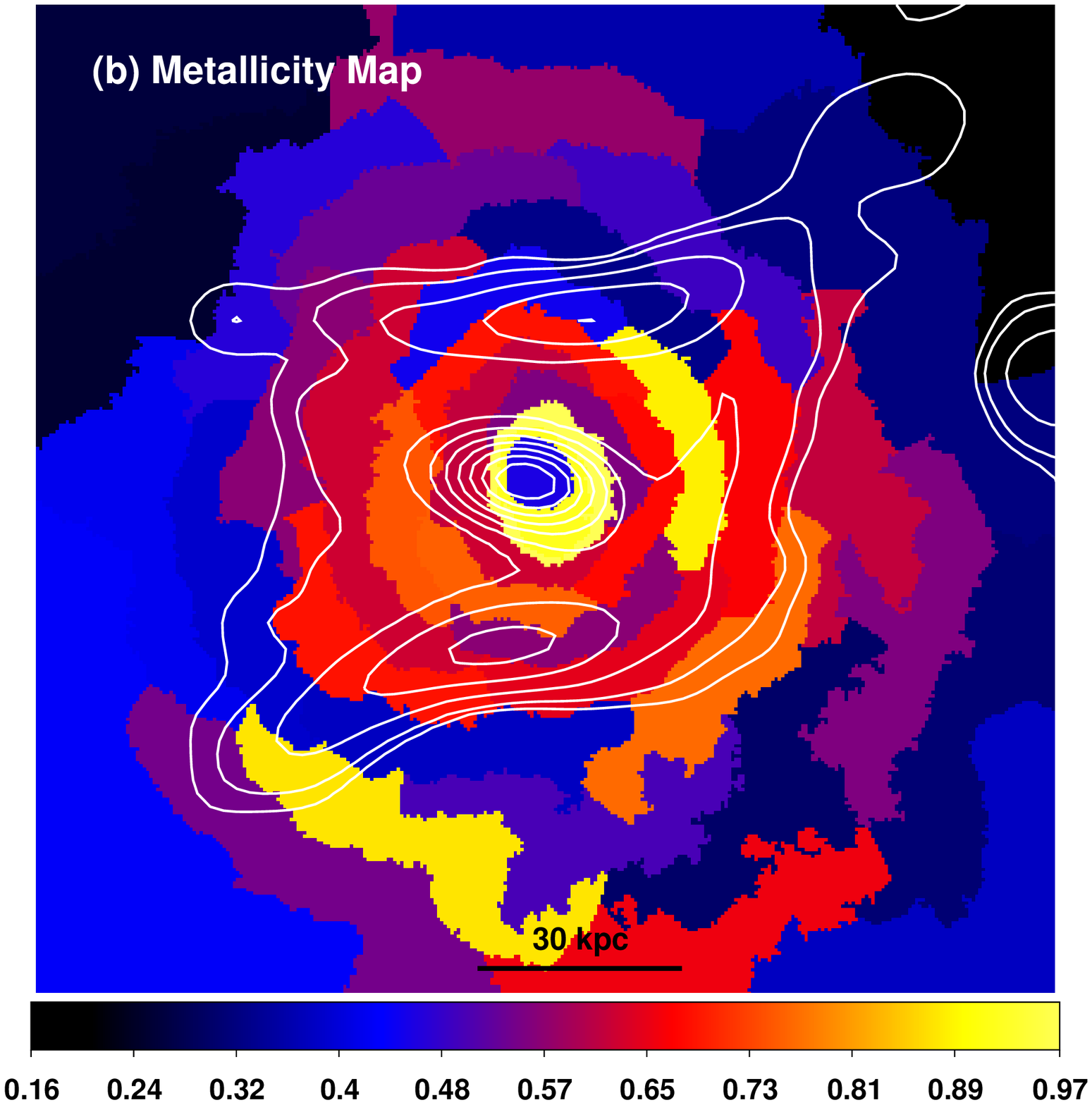}
}
\caption{Projected 2D maps of the thermodynamical parameters. ({\it left panel}) temperature (in units of keV) ({\it right panel}), metallicity (in units of \Zsun). These maps have been generated using the contour binning technique of \citet{2006MNRAS.371..829S}. VLA 1.45 GHz radio contours (white colour) are overlaid on to both the images to check their association with radio emission. The errors in the temperatures vary from 6\% in the central region to 10\% in the outer regions, while those in the metallicity varied from 25\% to 30\%.}
\label{contour_bin_map}
\end{figure*}


In view of the complex nature of the X-ray emission and merging of the sub-cluster for better understanding the ICM structure in this cluster we obtain 2D maps of the temperature and metallicity. The large number of counts detected in the combined data enabled us to derive these maps. For this, we construct contour bins of the spectral extraction from 0.5-7 keV \textit{Chandra} image following the procedure outlined \citep{2006MNRAS.371..829S}. Each of the spatial bin generated was ensured to have at least $\sim$3000 counts (S/N $\sim$ 50). The spectra and the responses for individual bins extracted from both the observations were then fitted simultaneously with an absorbed single temperature {\sc apec} model within {\sc sherpa}. As before, the absorbing column density and the redshift were fixed during the fit. Finally, 2D maps of temperature and metallicity were constructed using the \textit{paint\_output\_images} task and are shown in Figure~\ref{contour_bin_map}. 

Though the resulting temperature map (Figure~\ref{contour_bin_map} \textit{left panel}) confirms the general trends of increase, a clear asymmetry due to the complex nature of the ICM is evident. The metallicity map also exhibit asymmetries in the form of high-abundance arcs on the north-west and the south-west of the core. We expect the errors in the temperature measurement vary from 6\% in the core to about 10\% in the outer regions, while that for the metallicity varied from 25\% to 30\%. \cite{2013MNRAS.428...58R} propose that the observed structures in the ICM temperature could be due to the past AGN activity. However, it expects similar structures in the temperature and metallicity distribution on either sides of the centre, which is not the case with A2626. Therefore, they are likely be formed due to minor mergers. The southern arcs in the metallicity and temperature are found to be associated with the edge in the surface brightness distribution (Section~\ref{sec3.1_img}) and also show association with the VLA 1.45 GHz radio emission contours (Figure~\ref{contour_bin_map}). \cite{2018A&A...610A..89I} have proposed a possibility of merging of a subcluster with main cluster centred on A2626. The supporting evidences in favour of this merging are seen in the form of the asymmetric arcs, the cold fronts and the southern X-ray excess emission (Region 3 in Figure~\ref{fig3}). This merging probably leads to sloshing of the ICM and hence uplifts the high metallicity gas in the form of arcs.

\subsection{Discontinuities in the surface brightness}
\label{sec4.5_proffit}

\begin{figure*}
\centering
{
\includegraphics[scale=0.4]{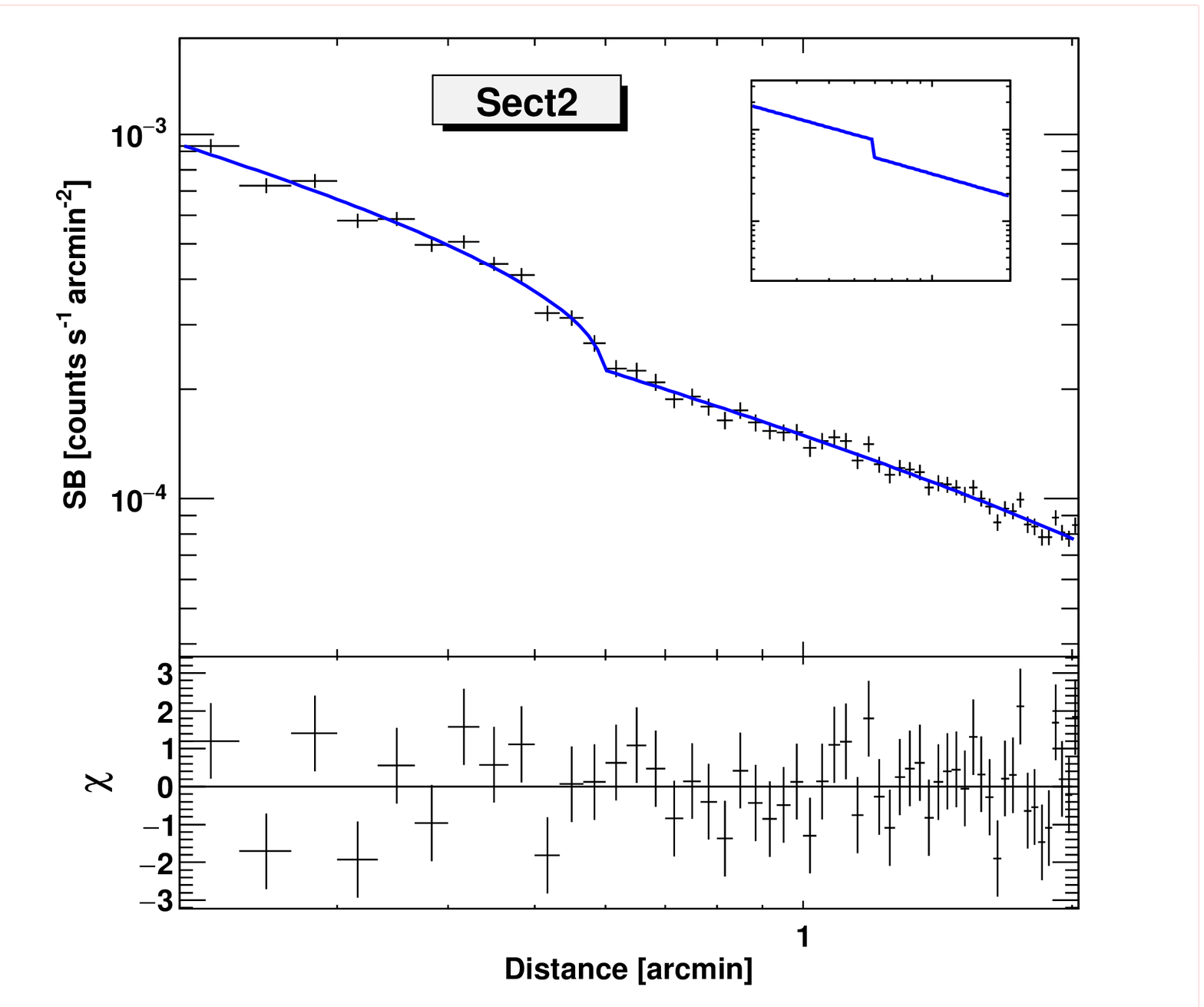}
\includegraphics[scale=0.4]{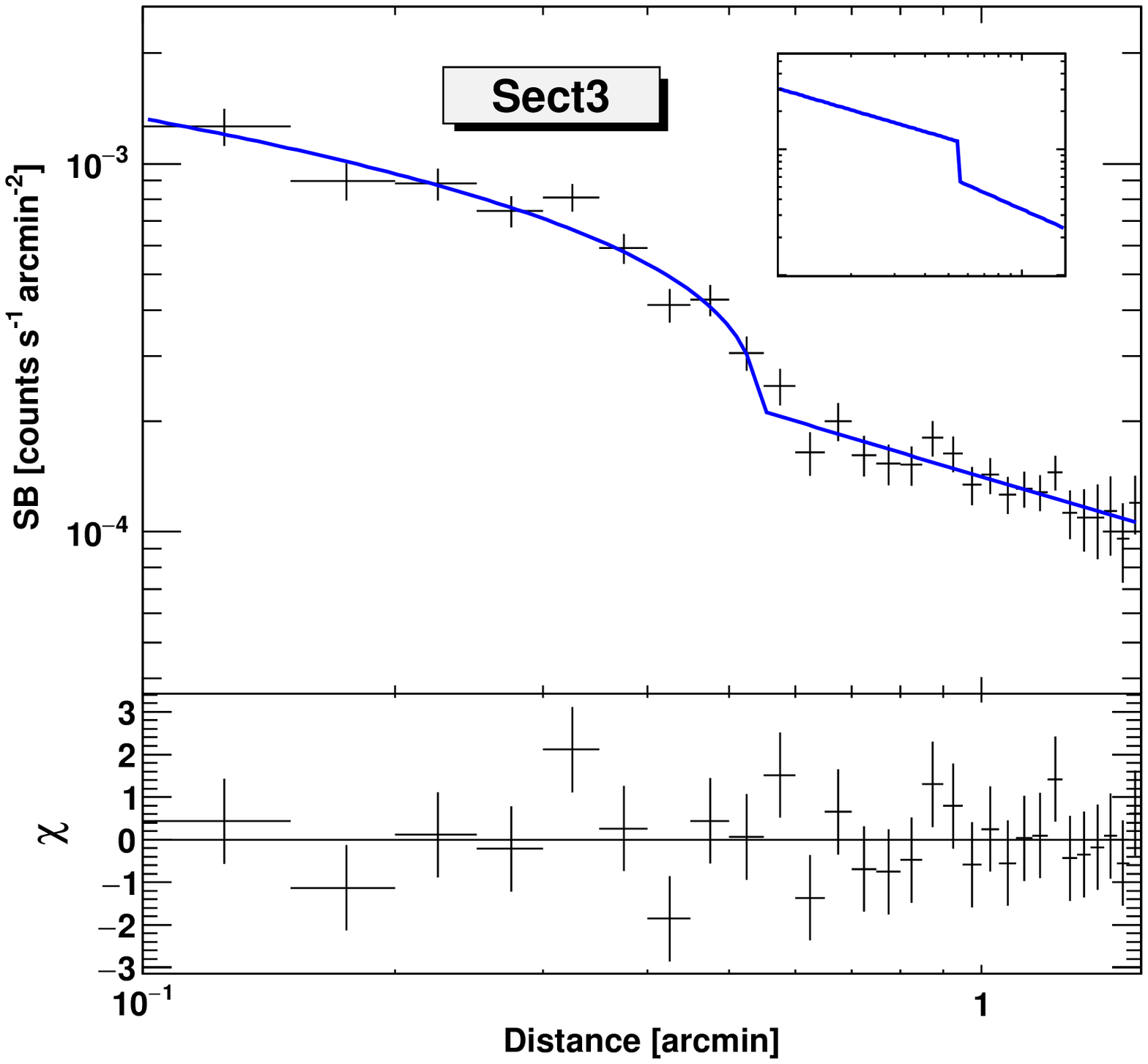}
}
\caption{X-ray surface brightness profiles in 0.5-3 keV energy band extracted from sectors in Sect2 ({\it left panel}) and Sect3 ({\it right panel}). These profiles were fitted with broken power-law density model shown in solid lines. Insets in each panel show corresponding 3D simulated gas density model. The bottom panels represent residuals due to fit.}
\label{fig6}
\end{figure*}


\begin{table*}
{\footnotesize
\caption{Parameters of the broken power-law (PROFFIT) fit to the surface brightness edge along sect2 and sect3. }
{\footnotesize {\hspace{2mm}} \begin{tabular}{|c|c|c|c|c|c|c|}
  \hline
  Angular Sectors  & $\alpha 1$     & $\alpha 2$ & r (kpc)     &  Norm N$_0$ (10$^{- 4}$)  & 
  C  & X$^2$/dof \\
  \hline
  280$^0$-350$^0$ (Sect2)  & 0.77$\pm$0.06  & 0.81$\pm$0.09  & 36$\pm$1  & 5.43$\pm$0.46  & 1.57$\pm$0.08  &  37.48/30 \\
  350$^0$-40$^0$ (Sect3)   & 0.57$\pm$0.11  & 0.86$\pm$0.08  & 33$\pm$1  & 8.20$\pm$1.37  & 2.06$\pm$0.44  & 21.28/22 \\
  \hline
  {
  } &  &  &  &  &  & 
\end{tabular}\label{c2tabX}}}
\end{table*}

Assuming that the edges in the surface brightness profiles along sectors Sect2 and Sect3 are  associated with the curved geometry of the cold regions in the temperature map Figure~\ref{contour_bin_map}, we investigate their relevance with the cold fronts. The cold fronts are the contact discontinuities in the surface brightness distribution due to the sloshing of the gas, where a drop in temperature and rise in gas density is expected, and are normally seen at the junctions of the merging clusters \citep{2000ApJ...541..542M,2001ApJ...551..160V}. For this, we extract the background subtracted 0.5-3 keV counts from sectors Sect2 and 3 after dividing them in to sectorial bins. The surface brightness along these sectors were then fitted with the deprojected broken power-law density model {\sc Proffit} (version 4.1)  developed by \cite{2011A&A...526A..79E} 
\begin{eqnarray}
n(r) = \begin{cases} C\,n_{\rm {0}} \left(\frac{r}{r_{\rm front}}\right)^{-\alpha1}\,, & \mbox{if } r \le r_{\rm front} \\ n_{\rm {0}} \left(\frac{r}{r_{\rm front}}\right)^{-\alpha2}\,, & \mbox{if } r > r_{\rm front} \end{cases} \,,
\end{eqnarray}
here, $n(r)$ gives the electron density at the projected distance $r$, $C = n_{e_2} / n_{e_1}$ the density compression factor at the edge, $\alpha_1$, $\alpha_2$ the power-law indices, and $r_{front}$ the radius at the putative cold front. This exercise provides with a precise way of investigating discontinuities in the surface brightness distributions. 

The extracted surface brightness profiles along with their best fit broken power-law models for Sect2 and Sect3 are shown in Figure~\ref{fig6} and the best-fit parameters are listed in Table~\ref{c2tabX}. This analysis exhibited discontinuities along both the sectors Sect 2 and 3, respectively, at 36 kpc and 33 kpc with the ICM density jumps equal to $1.57\pm0.08$ and $2.06\pm0.44$. These edges are due to the cold fronts and are supported by the small discontinuities in the temperature, metallicity, pressure, electron density and the entropy profiles (Figure~\ref{azmuth_spec}). The arcs in the temperature and metallicity maps and arcs in the radio emission (Figure~\ref{contour_bin_map}) also show their association with these fronts. In a parallel work using the recent data (ObsID 16136) \cite{2018A&A...610A..89I} also confirm the presence of cold fronts along both these sectors. However, the positions at which they find these cold fronts are slightly different, $\sim$29.5 kpc and 26.4 kpc along Sect 2 and 3, respectively. This is because the opening angels used by them for the sectorial extraction are different. They have noticed an inward temperature drops at both the locations in projected as well as de-projected temperature profiles, and are in agreement with our observations. They have also reported the association of both these cold fronts with the arcs in the diamond shaped radio emission and predict its origin due to the sloshing of the gas in south region. Thus, the arc shaped cold fronts are likely be formed due to sloshing of the gas at the interaction of the infalling sub-cluster with the main cluster (purple arcs in Figure~\ref{fig3} {\it left panel}). Similar inferences were also arrived at by \cite{2008ApJ...682..155W,2017MNRAS.466L..19K,2018A&A...610A..89I}. \cite{2008ApJ...682..155W}  witnessed a link between the excess X-ray emission due to IC 5337 on the west and the diffuse radio emission, providing another evidence in favour of the in-falling of the gas onto the centre of the main cluster.

\subsection{The nuclear sources}

\begin{figure*}
\centering
\includegraphics[scale=0.8]{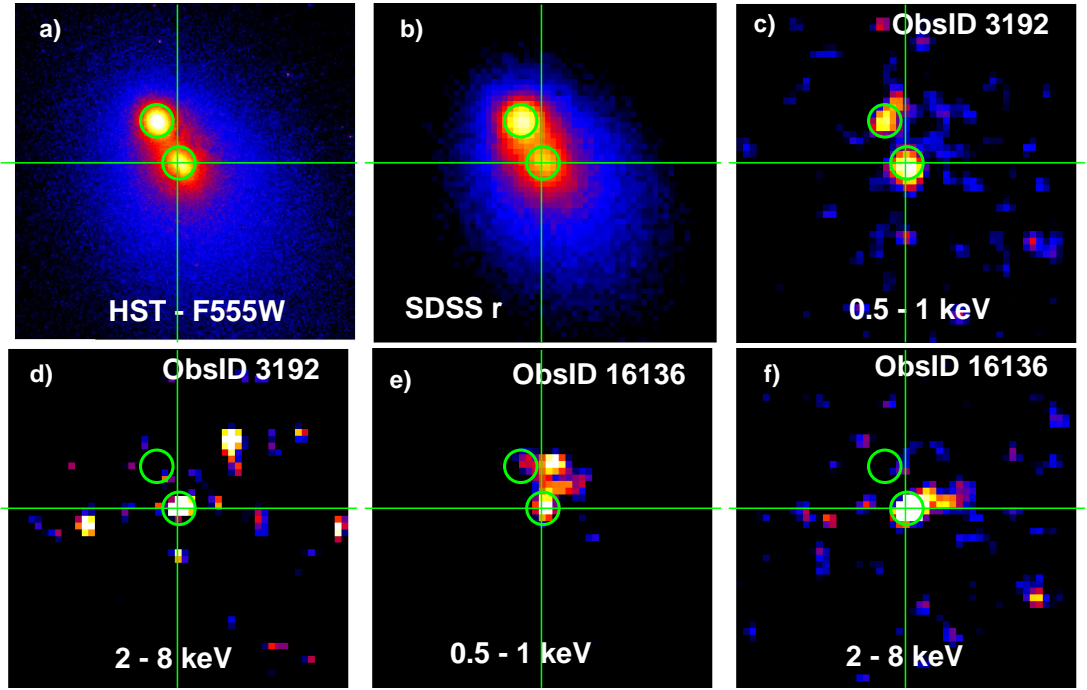}
\caption{Optical and X-ray imagery of central 25\arcsec x 25\arcsec of A2626 (a) HST F555W image, (b) SDSS r band image, (c), (d) \textit{Chandra} images using ObsId 3192. (e) and (f) \textit{Chandra} images using ObsId 16136. Each of the \textit{Chandra} image is divided in to two different energy bins the soft (0.5 - 1 keV) and the hard (2 - 8 keV) and are smoothed using 2 pixel wide Gaussian.  
}
\label{twosources}
\end{figure*}

\begin{table*}
\caption{Properties of the extracted counts from 1.5\arcsec\,regions centred on the two nuclear sources in Figure~\ref{twosources} for two different \textit{Chandra} ObsIDs.}
\begin{center}
\begin{tabular}{cccccccc}
\hline
& & & From event file & & & From Spectrum & \\\hline
ObsID		 &Regions & Soft Rate  	& Hard Rate   & Count Hardness      & Soft Rate  	  & Hard Rate      & Spectral Hardness \\
		 &        & ($10^{-3}$ ct/s)&($10^{-3}$ ct/s)& (Hard/Soft Rate) & ($10^{-3}$ ct/s)&($10^{-3}$ ct/s)& (Hard/Soft Rate)\\
\hline
3192   	 	 &NE-Source  & 1.27$\pm$0.23   & 0.25$\pm$0.10 & 0.20$\pm$0.08      & 1.16$\pm$0.27   & 0.026$\pm$0.158     & 0.01$\pm$0.09\\
                 &SW-Source  & 1.47$\pm$0.24   & 0.65$\pm$0.16 & 0.44$\pm$0.13  & 1.69$\pm$0.31   & 0.437$\pm$0.2     & 0.26$\pm$0.13\\
16136         	 &NE-Source  & 0.43$\pm$0.06   & 0.15$\pm$0.03 & 0.35$\pm$0.08  & 0.38$\pm$0.08   & 0.008$\pm$0.06     & 0.02$\pm$0.15\\
	         &SW-Source  & 0.55$\pm$0.07   & 0.72$\pm$0.08 & 1.30$\pm$0.22  	 & 0.53$\pm$0.09   & 0.59$\pm$0.09     & 1.13$\pm$0.26\\
\hline
\label{tab2}
\end{tabular}
\end{center}
\end{table*}

Central 20\arcsec$\times$20\arcsec\, region of A2626 imaged with HST F355W revealed clear detection of a pair of nuclear sources associated with the core of this cluster. With an objective to investigate their X-ray counterparts we examined the central 25\arcsec$\times$25\arcsec\, region of A2626 in both the \textit{Chandra} observations and found a pair of nuclear sources, the NE and the SW, separated by about 3.4~kpc, in agreement with those reported by \cite{2008ApJ...682..155W} using the old 24 ks data. We show the X-ray images of the central region of A2626 in two different energy bands (the soft 0.5--1 keV and hard 2--8 keV) for both the observations in Figure~\ref{twosources} ($c-e$). In the X--ray imaging, the SW source exhibits a clear detection in both the energy bands of both the observations. However, the NE source which emits most of its energy in the soft band of the old data, was barely detected in the recent deeper observation. Instead, we detect a soft component at about $2.''2$ towards the west of the NE source's optical position in the recent observation (Figure~\ref{twosources} $c$ and $e$). Given the 90\% uncertainty in the \chandra absolute astrometry corresponds to a shift as high as $0.8''$. Even if we assume 99\% uncertainty, the expected shift will not be more than 1.\arcsec4. Therefore, it is unlikely that the positional change of the NE source is due to the incorrect absolute astrometry of \chandra. Alternately, it could be attributed to the positional shift of this source over the span of 10 years.  However, this will require a velocity of $\sim 675\times{}c$ to move to the present position relative to its appearance in the old observation and is highly unrealistic. Also the fact that the NE source has not changed its position and brightness in  its corresponding optical observations (Figure~\ref{twosources} $a$, $b$) though were separated by six years, rules out this possibility. The bare detection of the NE source in the recent observation can also be explained by considering that the source has underwent into the spectral change or has faded away over the observational span. The separation of 10 years between the two observations provides enough time for a super massive black hole to undergo spectral change or fading \citep{2016MNRAS.457..389M,2018MNRAS.480.3898N,2018A&A...618A..83Z}.

To investigate this, we measured the count rates from the circular regions marked in Figure~\ref{twosources} and obtained the hardness of both the sources in each observation and are tabulated in columns 3 and 4 of Table~\ref{tab2}. A comparison of which revealed that the count rates in the soft band have decreased for both the sources. However, the hardness ratio of the NE source in both the observations has remained nearly same while that of the SW has increased by almost three times. The change in the hardness of an X-ray source could be attributed to the contamination layer on the blocking filter of the instrument involved. But it must be noted that such a contamination will affect identically for both the sources and hence change their hardness values similarly. Therefore, it is likely that the observed changes in the hardness of the SW source would be related to its intrinsic cause.

\begin{figure}
\centering
{
\includegraphics[scale=1]{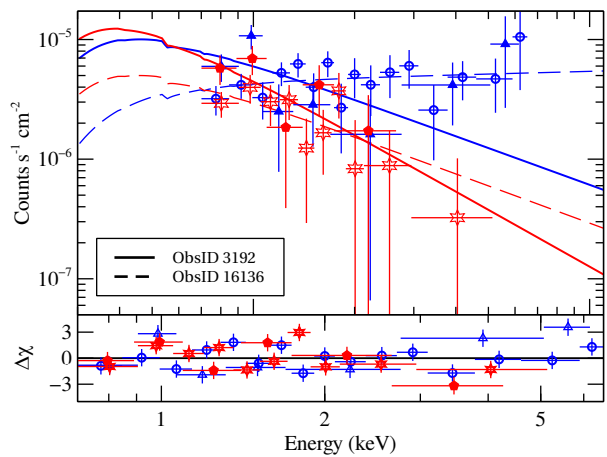}
}
\caption{The source spectra extracted from $1.5''$ circular regions centred on the NE (red) and SW (blue) nuclei for ObsID 3192 (filled symbols) and ObsID 16136 (open symbols). The best fitted absorbed power law model are also shown along with the $\Delta\chi$ variations (bottom panel). The spectra are fitted with $C-$statistics and are rebinned for plotting purpose.}
\label{spec}
\end{figure}

To confirm it further, we also compare their spectral properties in both the observations independently. For this we have extracted X-ray spectra from 1.\arcsec5 regions centred on each of the source for both the observations. Local background spectra just outside the source region were also extracted. We then generated the response matrices for each of the source in both the observations, which automatically takes care of the time dependent contamination, if any. Then we obtained the count rates in both the bands using \textit{XSPEC} and are shown in columns 6 and 7 of Table~\ref{tab2}. The resultant hardness ratios (column 8 in this table) estimated using the spectroscopic study are in--line with those derived earlier through the count rate statistics. We also tried to fit their spectra using a simple power law component modified by the Galactic absorption ($TBabs\times{}powerlaw$) in XSPEC. As the number of counts available for this fit were very few therefore we used the $C$-statistics during the spectral fit. All the parameters of the power law component were allowed to vary independently. This provided a reasonable fit with $C-$statistic of 352.7 for 360 degrees of freedom with the null hypothesis equal to $p_{null}$ = 0.64. The resultant spectrum is shown in Figure~\ref{spec}.  

The photon index ($\Gamma_{PL}$) for SW source changes from 3.02$^{+0.98}_{-0.72}$ to 1.92$^{+0.25}_{-0.24}$ between the two observations, which in turn imply that the source has moved into the hard state compared to the older observation. However, this is not the case for the NE source. The photon index for NE source was found to be 3.58$^{+1.10}_{-0.87}$ and 3.04$^{+0.60}_{-0.47}$ for the old and recent observations, respectively, and is consistent at 90\% confidence range. Similarly, the variation in the power law normalization was also within 90\% confidence level. 

The spectral analysis compliments our count rate statistics since the spectral hardness of the NE source remained almost constant in both the observations. The observed decrease in the count rates of the SW source could, therefore, be attributed to the fading of the AGN probably due to the variation in the mass accretion rate onto the central black hole \citep[e.g.,][]{2005Natur.433..604D,2006ApJS..163....1H}. Although simulation study expect the typical timescale of $\sim$ 10$^4$ years for such a change, there are several sources that showed fading or change in states (from type 1 to type 2) within few decades \citep{1999ApJ...519L.123A,2005ApJ...623L..93R,2014ApJ...796..134D,2015ApJ...800..144L,2015AJ....149..155K,2017ApJ...850..168K}. Another evidence in support of the fading of the AGN was provided by their associated outflows \citep{2018MNRAS.473.3525Z}. A massive outflow has been confirmed in the case of A2626 in MUSE observations \citep{June2017}. An alternative explanation for such a change was provided by the observational and simulation studies of interacting galaxies, where heavily obscured source due to the increased local absorption column density could also show a drop in its low energy photon flux \citep{2015ApJ...814..104K, 2017MNRAS.468.1273R}. However, in such situations the source would become harder, which is not true for the case of the NE source and hence can be ruled out. We tried to test this possibility by adding another absorption component to the power law during the fit but the poor statistics could not constrain the photon indices. The present data is not enough to arrive at a conclusive scenario for A2626, therefore, another deeper simultaneous covering the soft and the hard bands using sensitive instruments like XMM-Newton is called for exploring these possibilities in a systematic way.\\

\section[4]{Summary}
\label{sec4_summary}
This paper presents analysis of a combined 134 ks {\it Chandra} data of a peculiar galaxy cluster A2626 at a redshift of $z$=0.0553. This deeper dataset allowed us to investigate the substructures in the central region of this cluster in more details compared to those reported earlier using single shallow observations. This analysis confirms the earlier reported east cavity at 13.1 kpc and detects one more cavity at 38.9 kpc on the west of the X-ray centre of A2626 and were confirmed through a number of image processing techniques as well as the count statistics method. The average mechanical power injected by the AGN outburst was estimated to be ${\rm P_{cav} \sim 6.6 \times 10^{44}\, erg\, s^{-1}}$ and is roughly 29 times more than required to counter balance the cooling luminosity of ${\rm L_{cool} = 2.30 \pm 0.02 \times 10^{43} \ergs}$. The equivalence between the cavity power and the monochromatic 1.4 GHz radio power provided another evidence for the AGN heating of the ICM.

The SB profiles along the west and the south-west sectors exhibited edges, whose exact positions were measured by fitting them with the deprojected broken power-law density model at 33 kpc on the west and at 36 kpc on the south-west of the X-ray peak. Temperature drops were seen at both these edges, while the pressure profiles were almost continuous. We measure the gas density ratios of ${\rm n_{in}/n_{out} = 2.06\pm0.44}$ and 1.57$\pm$0.08, respectively, at these edges and are consistent with the typical values of cold fronts in galaxy clusters. These edges are also associated with the arcs in the temperature and metallicity maps and radio arcs support the sloshing of the gas in A2626.

This paper also reports the systematic study of a pair of nuclear sources using two independent observations separated by ten years. The NE source, which previously emitted most of its energy in the soft band, dramatically disappeared in the recent data. Instead an excess emission was seen at $2.2''$ on its west and required a line of sight velocity of $\sim$ $675\times{}c$ if is due to its movement. The count rate analysis and the spectral analysis confirms the dramatic change in the state of the SW source from type 1 to type 2 possibly due to the change in the mass accretion rate. No such change was noticed for the NE source. The low counts despite of deep observation could either be due to the increased absorption column or change in its spectral state as a result of the changed accretion rate. The former will make the spectrum harder, which is not seen in the spectral analysis. Another deep X-ray observation is called for disentangling these scenarios.

\section*{Acknowledgments}
{We thankfully acknowledge the sincere reading of the manuscript by the anonymous referee and also for the constructive and encouraging comments on the draft. This has helped us a lot to improve its quality. SKK gratefully acknowledges financial support by the Ministry of Minority Affairs, Govt. of India, under the Minority Fellowship Program. The data for this work has been obtained from the \chandra\,Data Archive, NASA/IPAC Extragalactic Database (NED), High Energy Astrophysics Science Archive Research Center (HEASARC). This work has made use of software packages CIAO and Sherpa provided by the \chandra X-ray Center. This work also used data from the NASA/ESA {\it Hubble Space Telescope}, at the Space Telescope Science Institute, which is operated by the Association of Universities for Research in Astronomy, Inc.}

\def\aj{AJ}%
\def\actaa{Acta Astron.}%
\def\araa{ARA\&A}%
\def\apj{ApJ}%
\def\apjl{ApJ}%
\def\apjs{ApJS}%
\def\ao{Appl.~Opt.}%
\def\apss{Ap\&SS}
\def\aap{A\&A}%
\def\aapr{A\&A~Rev.}%
\def\aaps{A\&AS}%
\def\azh{AZh}%
\def\baas{BAAS}%
\def\bac{Bull. astr. Inst. Czechosl.}%
\def\caa{Chinese Astron. Astrophys.}%
\def\cjaa{Chinese J. Astron. Astrophys.}%
\def\icarus{Icarus}%
\def\jcap{J. Cosmology Astropart. Phys.}%
\def\jrasc{JRASC}%
\def\mnras{MNRAS}%
\def\memras{MmRAS}%
\def\na{New A}%
\def\nar{New A Rev.}%
\def\pasa{PASA}%
\def\pra{Phys.~Rev.~A}%
\def\prb{Phys.~Rev.~B}%
\def\prc{Phys.~Rev.~C}%
\def\prd{Phys.~Rev.~D}%
\def\pre{Phys.~Rev.~E}%
\def\prl{Phys.~Rev.~Lett.}%
\def\pasp{PASP}%
\def\pasj{PASJ}%
\def\qjras{QJRAS}%
\def\rmxaa{Rev. Mexicana Astron. Astrofis.}%
\def\skytel{S\&T}%
\def\solphys{Sol.~Phys.}%
\def\sovast{Soviet~Ast.}%
\def\ssr{Space~Sci.~Rev.}%
\def\zap{ZAp}%
\def\nat{Nature}%
\def\iaucirc{IAU~Circ.}%
\def\aplett{Astrophys.~Lett.}%
\def\apspr{Astrophys.~Space~Phys.~Res.}%
\def\bain{Bull.~Astron.~Inst.~Netherlands}%
\def\fcp{Fund.~Cosmic~Phys.}%
\def\gca{Geochim.~Cosmochim.~Acta}%
\def\grl{Geophys.~Res.~Lett.}%
\def\jcp{J.~Chem.~Phys.}%
\def\jgr{J.~Geophys.~Res.}%
\def\jqsrt{J.~Quant.~Spec.~Radiat.~Transf.}%
\def\memsai{Mem.~Soc.~Astron.~Italiana}%
\def\nphysa{Nucl.~Phys.~A}%
\def\physrep{Phys.~Rep.}%
\def\physscr{Phys.~Scr}%
\def\planss{Planet.~Space~Sci.}%
\def\procspie{Proc.~SPIE}%
\bibliographystyle{mn.bst}
\bibliography{mybib.bib}
\end{document}